\documentclass[aps,prb,citeautoscript,twocolumn,superscriptaddress,footinbib,10pt]{revtex4-1} 
\usepackage[latin1]{inputenc}
\usepackage{amsmath,amssymb}
\usepackage{mathrsfs} 
\usepackage{graphicx,color,colortbl}
\usepackage{rotating,array,tabularx,booktabs}
\usepackage{varwidth,xcolor}
\usepackage{placeins}
\usepackage{siunitx}

\DeclareGraphicsExtensions{.pdf,.PDF,.png,.PNG}

\newcommand{\ZT}[1]{\textquotedblleft#1\textquotedblright}%

\newcommand{\CC}{\boldsymbol{\mathrm{C}}}%
\newcommand{\DD}{\boldsymbol{\mathcal{D}}}%
\newcommand{\HH}{\boldsymbol{\mathcal{H}}}%
\newcommand{\KK}{\boldsymbol{\mathrm{K}}}%
\newcommand{\MM}{\boldsymbol{\mathrm{A}}}%
\newcommand{\OO}{\boldsymbol{\Omega}}%
\newcommand{\RR}{\boldsymbol{\mathcal{R}}}%

\begin{document}

\title{Hydrodynamic resistance matrices of colloidal particles with various shapes}
\author{Johannes Vo\ss{}}
\affiliation{Institut f\"{u}r Theoretische Physik, Westf\"alische Wilhelms-Universit\"{a}t M\"{u}nster, D-48149 M\"{u}nster, Germany}
\affiliation{Center for Soft Nanoscience, Westf\"alische Wilhelms-Universit\"{a}t M\"{u}nster, D-48149 M\"{u}nster, Germany}

\author{Raphael Wittkowski}
\email[Corresponding author: ]{raphael.wittkowski@uni-muenster.de}
\affiliation{Institut f\"{u}r Theoretische Physik, Westf\"alische Wilhelms-Universit\"{a}t M\"{u}nster, D-48149 M\"{u}nster, Germany}
\affiliation{Center for Soft Nanoscience, Westf\"alische Wilhelms-Universit\"{a}t M\"{u}nster, D-48149 M\"{u}nster, Germany}
\affiliation{Center for Nonlinear Science, Westf\"alische Wilhelms-Universit\"{a}t M\"{u}nster, D-48149 M\"{u}nster, Germany}

\begin{abstract}
The hydrodynamic resistance matrix is an important quantity for describing the dynamics of colloidal particles. This matrix encodes the shape- and size-dependent hydrodynamic properties of a particle suspended in a simple liquid at low Reynolds number and determines the particle's diffusion tensor. For this reason, the hydrodynamic resistance matrix is typically needed when modeling the motion of free purely Brownian, externally driven, or self-propelled colloidal particles or the behavior of dilute suspensions of such particles on the basis of Langevin equations, Smoluchowski equations, classical dynamical density functional theory, or other appropriate methods. So far, however, the hydrodynamic resistance matrix was available only for a few particle shapes. In this article, we therefore present the hydrodynamic resistance matrices for various particle shapes that are relevant for current research, including apolar and polar as well as convex and partially concave shapes. The elements of the hydrodynamic resistance matrices are given as functions of shape parameters like the aspect ratio of the corresponding particle so that the results apply not only to discrete but instead to continuous sets of particle shapes. This work shall stimulate and support future studies on colloidal particles with anisometric shapes.  
\end{abstract}

\maketitle

\section{Introduction}
During the last three decades an enhanced interest in anisometric colloidal particles has led to large progress in the synthesis of particles with predefined shapes  \cite{YinA2005,BurdaCNES2005,TaoHY2008,ChampionKM2007,SacannaP2011,KuijkvBI2011}. When describing the dynamics of a rigid colloidal particle in a simple liquid at low Reynolds number theoretically, the particle's hydrodynamic resistance matrix \cite{Brenner1967,HappelB1991} (also called \ZT{hydrodynamic friction tensor} \cite{KraftWtHEPL2013}) is the main quantity that is needed as input. 
It depends only on the shape and size of the particle and -- for a given dynamic viscosity of the liquid -- determines the translational and rotational hydrodynamic drag that acts on the particle when it moves relative to the liquid. 
 
Through a generalized Stokes-Einstein relation, the hydrodynamic resistance matrix provides the diffusion tensor of a colloidal particle. This matrix is therefore highly relevant even when the Brownian motion of a free particle shall be described. Likewise, it is needed for specifying the Langevin equations of colloidal particles driven by external forces or torques, which allow, e.g., to calculate the velocity of a sedimenting particle, as well as of self-propelled (also called \ZT{active} \cite{BechingerdLLRVV2016}) colloidal particles \cite{WittkowskiL2012,KuemmeltHWBEVLB2013,KuemmeltHWTBEVLB2014,tenHagenKWTLB2014}.
Also when applying Smoluchowski equations for the dynamics of an individual particle or of a dilute suspension of colloidal particles \cite{Dhont1996}, the hydrodynamic resistance matrix is a necessary input quantity. Furthermore, this matrix is important in the context of classical dynamical density functional theory for colloidal particles with anisometric shapes \cite{WittkowskiL2011}.    

While the size-dependence of the hydrodynamic resistance matrix is given by a simple scaling law (see below), the shape-dependence is highly nontrivial. This complicates the determination of the hydrodynamic resistance matrix for a particular particle shape. 
The hydrodynamic resistance matrix is analytically known for spheres and spheroids \cite{Perrin1936,Simha1940}. For a few other particle shapes like cylinders and dumbbells \cite{HappelB1991}, at least approximate analytical expressions are available. A method that allows to derive analytical approximations for the hydrodynamic resistance matrix of a slender particle is slender-body theory \cite{Batchelor1970}. Examples of its typical applications are nanotubes \cite{FaganBCH2008}, nanowires \cite{AgarwalLRYLG2005}, slender kinked particles \cite{tenHagenWTKBL2015}, and fibers \cite{ClagueP1997,SwitzerK2003}. 

A numerical calculation of the hydrodynamic resistance matrix for a specific particle shape is in general possible, but it can be computationally very expensive.
This applies especially to the straightforward route via numerically solving the Stokes equation that describes the flow and pressure fields around the considered particle when it moves through an unbounded liquid at low Reynolds number \cite{HappelB1991}. 
Much faster, but usually not exact, are bead-model-based calculations \cite{SwansonTdH1978,GarciadelaTorreB1981,CarrascoGdlT1999,GarciadelaTorreC2002,Hansen2004,BetBDvR2017} as implemented in the program suite \texttt{HYDRO} \cite{GarciadelaTorreB1981,CarrascoGdlT1999,GarciadelaTorreC2002}. The main idea of such calculations is to represent the prescribed particle shape by a set of spheres and to calculate the pairwise hydrodynamic interactions between the spheres (i.e., the \ZT{beads}) analytically. This method is very useful for particles such as molecules consisting of nonoverlapping spheres, where it can provide exact results for the hydrodynamic resistance matrix. 
Currently, numerical results for the hydrodynamic resistance matrix or at least a few of its elements are known for only some particle shapes.
These include cylinders \cite{SwansonTdH1978,TiradoGdlT1979,Hansen2004,PassowtHLW2015}, spherocylinders \cite{PassowtHLW2015}, spindle shapes \cite{PassowtHLW2015}, double cones \cite{PassowtHLW2015}, Platonic solids \cite{GarciadelaTorreREO2007,BetBDvR2017}, red-blood-cell shapes \cite{MauerPPGF2017}, hollow spherical caps \cite{SwansonTdH1978}, microwedges \cite{KaiserPStHLA2014}, dumbbell shapes \cite{SwansonTdH1978,CarrascoGdlT1999}, chains of spheres \cite{JeffreyO1984,CarrascoGdlT1999,CarlsonJFCSW2006,GarciadelaTorreREO2007}, three-body swimmers consisting of Platonic solids \cite{BetBDvR2017},
a microswimmer that propels itself by a rotating helical flagellum \cite{BetBDvR2017}, oligomers \cite{CarrascoGdlT1999,GarciadelaTorreREO2007}, and macromolecules \cite{BernalGarciaGdlT1980,GarciadelaTorreB1981,MatulisBBL1999,GarciadelaTorreHC2000a,GarciadelaTorreHC2000b,GarciadelaTorreC2002,GarciadelaTorreREO2007}. 

Besides direct numerical calculations, experimental data from observing and analyzing a particle's orientation-resolved trajectories can be used to determine its hydrodynamic resistance matrix \cite{KraftWtHEPL2013,FungM2013}.
However, the shapes for which this matrix is known constitute only a very small amount of the particle shapes that can be synthesized and that are relevant for research.
Even for many rather symmetric basic shapes like spherocylinders and cones, the hydrodynamic resistance matrix is available only in the case of particular aspect ratios of the shape or not at all. 

Therefore, in this article, we present the hydrodynamic resistance matrices for various basic particle shapes. For all these shapes, we varied at least one shape parameter so that the elements of the matrices are given as functions of these parameters. 
The particle shapes we consider include apolar (i.e., with a head-tail symmetry) and polar (i.e., with a broken head-tail symmetry) as well as convex and partially concave shapes. 
In particular, they are right circular cylinders with plane ends, rectangular cuboids with quadratic cross sections, right circular cylinders with concave and convex spherical ends, right circular cylinders with concave and convex conical ends, hollow and full half balls, hollow and full right circular cones, as well as double-cup shapes with symmetry group $\mathrm{C_{2h}}$ consisting of two hollow or full half balls or right circular cones. We varied the aspect ratios of these particles, the curvature and thus height of the spherical ends, the length of the conical ends, the wall thickness of the hollow particles, and the width of the contact area of the particles that constitute the double-cup particles. 
The apolar rodlike shapes belong to the most frequently considered anisometric particles, the polar shapes have gained high relevance in the context of self-acoustophoretic particles \cite{WangCHM2012,GarciaGradillaEtAl2013}, and also shapes like that of the double-cup particles that have one plane of symmetry and are chiral with respect to that plane are of interest in current research on soft-matter systems \cite{ZerroukiBPCB2008,SokolovAGA2010,MijalkovV2013,SchamelPGMMF2013,KuemmeltHWBEVLB2013,KuemmeltHWTBEVLB2014,tenHagenKWTLB2014,YuanMSTS2018}. 
Our results rely on quite accurate direct numerical solutions of the Stokes equation and shall support future studies on anisometric colloidal particles.

The article is organized as follows: In Sec.\ \ref{methods}, we present background information on the hydrodynamic resistance matrix and the numerical methods we used for calculating it. The results of our calculations for various particle shapes are presented and discussed in Sec.~\ref{results}. Finally, we conclude in Sec.\ \ref{conclusions}.

\section{\label{methods}Methods}
The hydrodynamic resistance matrix $\HH$ is $6\!\times\!6$-dimensional and symmetric. When a particle in an unbounded liquid at low Reynolds number moves with translational velocity $\vec{v}$ and angular velocity $\vec{\omega}$ relative to the liquid, $\HH$ allows to determine the hydrodynamic drag force $\vec{F}$ and torque $\vec{T}$ acting on the particle. 
The relation between the force-torque vector $\vec{\mathfrak{F}}=(\vec{F},\vec{T})^{\mathrm{T}}$ and the translational-angular velocity vector $\vec{\mathfrak{v}}=(\vec{v},\vec{\omega})^{\mathrm{T}}$ is given by \cite{HappelB1991,WittkowskiL2012}
\begin{equation}
\vec{\mathfrak{F}}=-\eta\,\HH\,\vec{\mathfrak{v}} 
\end{equation}
with the liquid's dynamic (or \ZT{shear}) viscosity $\eta$. From the hydrodynamic resistance matrix of a particle, its $6\!\times\!6$-dimensional short-time diffusion tensor $\DD$ follows by the generalized Stokes-Einstein relation \cite{WittkowskiL2012} 
\begin{equation}
\DD=\frac{1}{\beta\eta}\,\RR^{-1}\,\HH^{-1}\,\RR \;.
\end{equation}
Here, $\beta=1/(k_{\mathrm{B}}T)$ denotes the inverse thermal energy (also called \ZT{thermodynamic beta}) with the Boltzmann constant $k_{\mathrm{B}}$ and the absolute temperature $T$. Furthermore, $\RR$ is a rotation matrix that depends on the orientation of the particle. This matrix maps from the laboratory frame to the particle-fixed coordinate system that has been used when calculating the hydrodynamic resistance matrix $\HH$ of the particle. Besides the position of the origin and the orientation of the particle-fixed coordinate system, the size and shape of the particle affect the elements of $\HH$.  

Taking its symmetry into account, the hydrodynamic resistance matrix $\HH$ can be written as the block matrix \cite{Brenner1967,WittkowskiL2012}
\begin{equation}
\HH=
\begin{pmatrix}
\KK & \CC^{\mathrm{T}}_{\mathrm{S}} \\
\CC_{\mathrm{S}} & \OO_{\mathrm{S}} 
\end{pmatrix} .
\label{eq:H}%
\end{equation}
The $3\!\times\!3$-dimensional submatrices $\KK$, $\CC_{\mathrm{S}}$, and $\OO_{\mathrm{S}}$ correspond to translation, translational-rotational coupling, and rotation, respectively.
While the translation tensor $\KK$ and the rotation tensor $\OO_{\mathrm{S}}$ are symmetric, the coupling tensor $\CC_{\mathrm{S}}$ is in general not symmetric. 
Changing the origin of the particle-fixed coordinate system has an effect on $\CC_{\mathrm{S}}$ and $\OO_{\mathrm{S}}$, but not on $\KK$. Therefore, this reference point is denoted by the symbol \ZT{$\mathrm{S}$} in $\CC_{\mathrm{S}}$ and $\OO_{\mathrm{S}}$. Throughout this article, we use the center of mass of a particle as its reference point $\mathrm{S}$. 
When $\HH$ is known for a certain reference point $\mathrm{S}$, the matrix corresponding to a different reference point $\mathrm{P}$ can be obtained by a simple transformation. 
For this transformation, the submatrices $\CC_{\mathrm{S}}=(C_{\mathrm{S},ij})_{i,j=1,2,3}$ and $\OO_{\mathrm{S}}=(\Omega_{\mathrm{S},ij})_{i,j=1,2,3}$ corresponding to $\mathrm{S}$ need to be replaced by the submatrices $\CC_{\mathrm{P}}=(C_{\mathrm{P},ij})_{i,j=1,2,3}$ and $\OO_{\mathrm{P}}=(\Omega_{\mathrm{P},ij})_{i,j=1,2,3}$ corresponding to $\mathrm{P}$. The latter matrices are given by \cite{HappelB1991} 
{\allowdisplaybreaks\begin{align}%
\begin{split}%
C_{\mathrm{P},ij} &= C_{\mathrm{S},ij} - \sum^{3}_{k,l=1} \varepsilon_{ikl} r_{\mathrm{SP},k} K_{lj} \;, 
\end{split}\\%
\begin{split}%
\Omega_{\mathrm{P},ij} &= \Omega_{\mathrm{S},ij}  
- \sum^{3}_{k,l=1} \varepsilon_{ikl} \sum^{3}_{m,n=1} \varepsilon_{jmn} r_{\mathrm{SP},k} K_{lm} r_{\mathrm{SP},n} \\
&\quad\,+ \sum^{3}_{k,l=1} \varepsilon_{jkl} C_{\mathrm{S},ik} r_{\mathrm{SP},l} - \sum^{3}_{k,l=1} \varepsilon_{ikl} r_{\mathrm{SP},k} C_{\mathrm{S},jl} 
\end{split}%
\end{align}}%
with the Levi-Civita symbol $\varepsilon_{ijk}$, the vector $\vec{r}_{\mathrm{SP}}=(r_{\mathrm{SP},1},r_{\mathrm{SP},2},r_{\mathrm{SP},3})^{\mathrm{T}}$ pointing from $\mathrm{S}$ to $\mathrm{P}$, and the unchanged translation tensor $\KK=(K_{ij})_{i,j=1,2,3}$.

As a consequence of its symmetry, the matrix $\HH$ has up to $21$ independent elements. Symmetries of the particle shape can reduce the number of independent elements of $\HH$ \cite{HappelB1991}. 
When $\vec{x}$ is a vector-valued variable that describes a position in three-dimensional space, the shape of a particle can be defined by an implicit function $f(\vec{x})=0$. 
A symmetry transformation given by a rotation, reflection, or combination of both that maps the particle shape and the reference point onto themselves can be represented by an orthogonal matrix $\MM$ with the property $f(\MM\vec{x})=0$. Such a symmetry property leads to the conditions \cite{HappelB1991}
{\allowdisplaybreaks\begin{gather}%
\MM^{-1}\KK\MM = \KK \;, \\
\MM^{-1}\CC_{\mathrm{S}}\MM = \det(\MM)\CC_{\mathrm{S}} \;, \\
\MM^{-1}\OO_{\mathrm{S}}\MM = \OO_{\mathrm{S}}  
\end{gather}}%
for the submatrices $\KK$, $\CC_{\mathrm{S}}$, and $\OO_{\mathrm{S}}$, where $\det(\MM)$ denotes the determinant of the transformation matrix $\MM$. 
When a symmetry transformation also maps axes or planes of the particle-fixed coordinate system onto themselves, typical consequences of the symmetry properties of the particle shape are that some elements of the corresponding hydrodynamic resistance matrix $\HH$ are zero, equal, or additive inverses of each other. 
This means that for particle shapes with symmetry properties the structure of $\HH$ can be simplified by choosing the position of the origin and the orientation of the particle-fixed coordinate system appropriately. 
Since the particle shapes we study in the following have symmetries, we use this feature of $\HH$ to minimize the number of its elements that have different absolute values. 

The dependence of the hydrodynamic resistance matrix $\HH$ on the particle size can be described by the simple scaling relations \cite{HappelB1991} 
\begin{equation}
K_{ij}\propto l \;, \qquad C_{\mathrm{S},ij}\propto l^{2} \;, \qquad \Omega_{\mathrm{S},ij}\propto l^{3}
\end{equation}
with the length scale $l$. In contrast, the dependence of $\HH$ on the shape of the particle is complicated.  
To obtain $\HH$ for a particular particle shape, we followed the straightforward route described in Ref.\ \onlinecite{HappelB1991}. It includes calculating the Stokes flow around the considered particle when it moves through a liquid that is quiescent at large distance from the particle and solving the appropriate integral equations that yield the elements of $\HH$.
For numerically solving the Stokes equation in a complex three-dimensional domain, we used the finite element method as implemented in the computing platform \texttt{FEniCS} \cite{Kirby2004,KirbyL2006,AlnaesLMSL2009,LoggW2010,OlgaardW2010,AlnaesLORW2012,LoggMW2012,AlnaesEtAl2015}. 
By setting the corresponding options in \texttt{FEniCS}, we chose the minimal residual method \texttt{minres} as solution method for linear systems of equations and the algebraic multigrid preconditioner \texttt{amg} to precondition the linear systems \cite{LoggMW2012}.
We simulated a particle in the middle of a cubic box with edge length $1000\sigma$, where $\sigma$ was the diameter or width of the particle. 
At the boundary of the simulation box and at the surface of the particle, we prescribed no-slip conditions. 
The simulation domain, i.e., the space inside the box except for the particle, was discretized by an unstructured tetrahedral mesh created with the mesh generator \texttt{Gmsh} \cite{GeuzaineR2009}.
This mesh was particularly fine close to the particle. Depending on the particle shape, the mesh included from some thousands to some hundred thousands triangles on the particle surface and a few million tetrahedra outside of the particle (see the Appendix for details). 
 
We calculated the hydrodynamic resistance matrix $\HH$ for all considered particle shapes, varying one or two shape parameters like the aspect ratio of the particle in small steps over certain intervals.
To express the dependence of the elements of $\HH$ on the shape parameters by approximate analytical functions, we fitted third-order polynomials with the shape parameters as variables to the simulation results. The determined functions are best-fit functions in the least-squares sense. By evaluating these fit functions for particular values of the corresponding shape parameters, values for the elements of $\HH$ that closely match, interpolate, or extrapolate the simulation results, are obtained. 
In the case of one varied shape parameter, we used the polynomial 
\begin{equation}
p_{1}(x) = a_{0} + a_{1}x + a_{2}x^{2} + a_{3}x^{3} 
\label{eq:pI}%
\end{equation}
with the fit coefficients $a_{0},\dotsc,a_{3}$. When two shape parameters were varied, the chosen polynomial was  
\begin{equation}
\begin{split}%
p_{2}(x,y) &= a_{0} + a_{1}x + a_{2}y + a_{3}x^{2} + a_{4}xy + a_{5}y^{2} \\
&\quad\,+ a_{6}x^{3} + a_{7}x^{2}y + a_{8}xy^{2} + a_{9}y^{3} 
\end{split}%
\label{eq:pII} %
\end{equation}
with the fit coefficients $a_{0},\dotsc,a_{9}$. To assess the accuracy of the best-fit functions, we calculated the root-mean-square deviation of each fit function from the simulation data. 

In the next section, the simulation results for the hydrodynamic resistance matrix are compared to our fit functions and to some available approximate analytical results from the literature. For further comparison, we calculated approximate numerical results for a particle by bead-model simulations with the software \texttt{HYDROSUB} \cite{GarciadelaTorreB1981,CarrascoGdlT1999,GarciadelaTorreC2002} from the program suite \texttt{HYDRO}.

\section{\label{results}Results and discussion}

\subsection{Right circular cylinder with plane ends and rectangular cuboid with quadratic cross section}
\begin{figure}[htb]
\centering
\includegraphics[width=\linewidth]{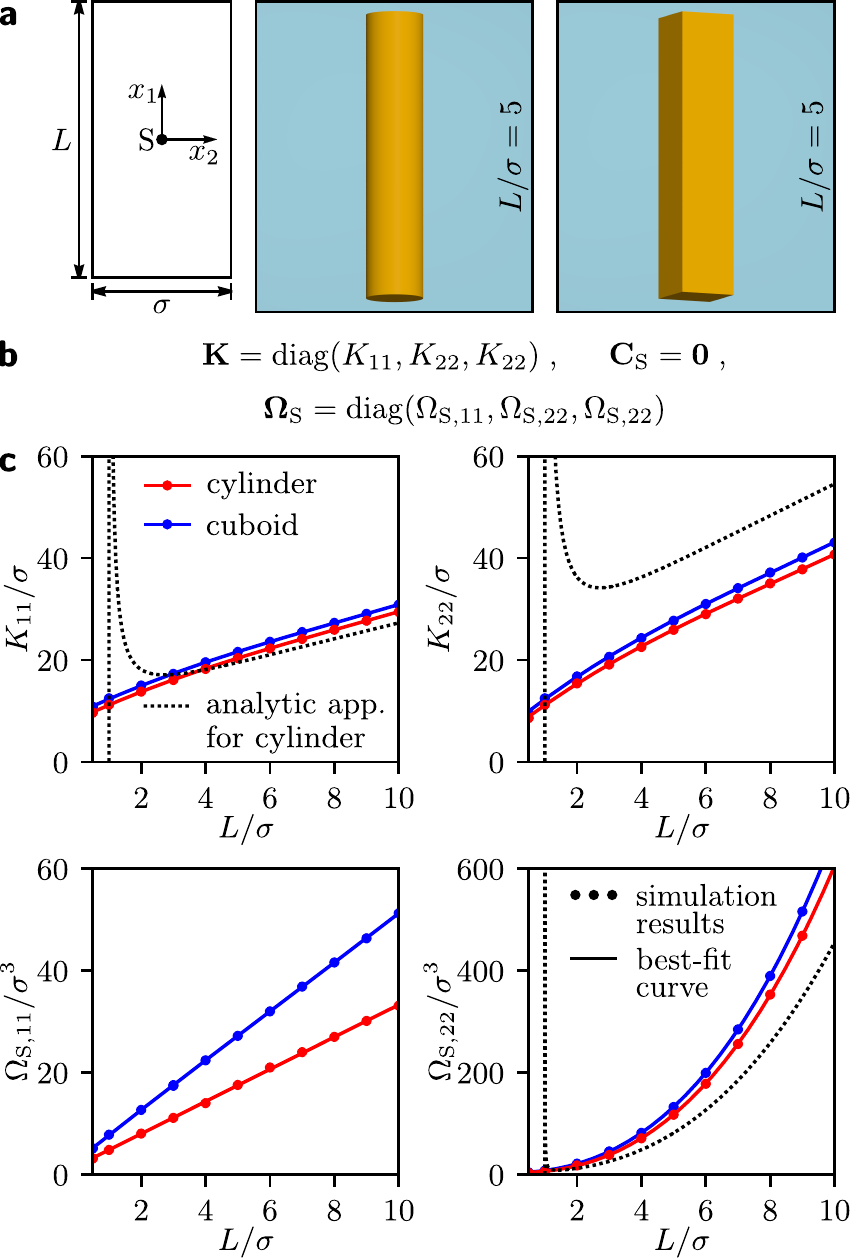}%
\caption{\label{fig:cylinder_cuboid}Results for the hydrodynamic resistance matrices of a right circular cylinder with plane ends and a rectangular cuboid with quadratic cross section. (a) Sketch showing a particle with diameter (for the cylinder) or width (for the cuboid) $\sigma$ and length $L$, its center of mass $\mathrm{S}$ as reference point, its orientation relative to the chosen right-handed coordinate system, and three-dimensional visualizations of the particles. (b) The structure of the hydrodynamic resistance submatrices $\mathbf{K}$, $\mathbf{C}_{\mathrm{S}}$, and $\mathbf{\Omega}_{\mathrm{S}}$ for the chosen reference point and orientation of the particles. (c) Simulation results for the nonvanishing elements of $\mathbf{K}$, $\mathbf{C}_{\mathrm{S}}$, and $\mathbf{\Omega}_{\mathrm{S}}$ as functions of $L/\sigma$ as well as the corresponding best-fit curves given by $p_{1}(L/\sigma)$ with the coefficients from Tabs.\ \ref{tab:cylinder} (for cylinder) and \ref{tab:cuboid} (for cuboid); The approximate analytic results \eqref{eq:H_cylinder_approx} for a cylinder are shown for comparison.}
\end{figure}

We start with a right circular cylinder with plane ends and a rectangular cuboid with quadratic cross section (see Fig.\ \ref{fig:cylinder_cuboid}a).
The cylinder has rotational symmetry about the $x_{1}$ axis and reflection symmetry with respect to the $x_{2}$-$x_{3}$ plane. Its shape is therefore \ZT{apolar}.
The cuboid has three pairwise perpendicular planes of symmetry, which are the three coordinate planes here. In addition, it has discrete rotational symmetry of the fourth order with respect to the $x_{1}$ axis. As a consequence of these symmetries, the hydrodynamic resistance matrix $\HH$ is diagonal and has only four independent nonzero elements for both particle shapes. These nonzero elements are $K_{11}$, $K_{22}$, $\Omega_{\mathrm{S},11}$, and $\Omega_{\mathrm{S},22}$ (see Fig.\ \ref{fig:cylinder_cuboid}b). This means that there is no translational-rotational coupling.    
Our direct simulation results for the nonzero elements of the hydrodynamic resistance matrices of the particles are shown as functions of $L/\sigma\in [0.5,10]$, where $L$ is the length and $\sigma$ is the diameter or width of a particle (see Fig.\ \ref{fig:cylinder_cuboid}c). 
The corresponding best-fit curves are given by the polynomial \eqref{eq:pI} with the best-fit values for the coefficients of the polynomial that are given in Tabs.\ \ref{tab:cylinder} and \ref{tab:cuboid} in the Appendix for the cylinder and cuboid, respectively. 
One can see that the direct simulation results and best-fit curves agree very well. 
For a given particle diameter or width $\sigma$, all nonzero elements of $\HH$ increase with $L/\sigma$. This is reasonable, since with $L$ both the surface of the particle and thus the hydrodynamic drag grow. The curves for cylinder and cuboid are qualitatively similar. However, for the cuboid they are always above and steeper than those for the cylinder.
This is consistent with the fact that for given $L$ and $\sigma$, the surface of the cuboid is a factor $4/\pi$ greater than the surface of the cylinder.
For a thin and long cylinder, there are analytic approximations for $K_{11}$, $K_{22}$, and $\Omega_{\mathrm{S},22}$ given by \cite{Dhont1996}
\begin{equation}
K_{11} \approx \frac{2 \pi L}{\ln\!\big(\frac{L}{\sigma}\big)} \;, \quad
K_{22} \approx \frac{4 \pi L}{\ln\!\big(\frac{L}{\sigma}\big)} \;, \quad 
\Omega_{\mathrm{S},22} \approx \frac{\pi L^3}{3\ln\!\big(\frac{L}{\sigma}\big)} \;. 
\label{eq:H_cylinder_approx}%
\end{equation}
Figure \ref{fig:cylinder_cuboid}c shows also curves corresponding to these analytic approximations. Their overall agreement with the simulation results is poor. For $L/\sigma\lesssim 2$, the analytic approximations are not applicable. There are strong deviations from the simulation data and even a divergence at $L/\sigma=1$. 
The analytic approximations are better for $L/\sigma\gtrsim 2$. In the case of $K_{11}$, the agreement is quite good. As opposed to this, the analytic approximation for $K_{22}$ leads to strongly too large values. In the case of $\Omega_{\mathrm{S},22}$, the agreement is good for $L/\sigma\approx 2$, but the analytic results are increasingly too small for growing $L/\sigma$.
For further comparison, we calculated values for the elements $K_{11}$, $K_{22}$, $\Omega_{\mathrm{S},11}$, and $\Omega_{\mathrm{S},22}$ with the software \texttt{HYDROSUB}. 
Assuming a cylinder with $L/\sigma=10$, the calculations led to  
$K_{11}/\sigma=\SI{29,51}{}$, $K_{22}/\sigma=\SI{40,57}{}$, $\Omega_{\mathrm{S},11}/\sigma^{3}=\SI{34,37}{}$, and $\Omega_{\mathrm{S},22}/\sigma^{3}=\SI{575,58}{}$.
The corresponding simulation results are $K_{11}/\sigma=\SI{29,46}{}$, $K_{22}/\sigma=\SI{40,73}{}$, $\Omega_{\mathrm{S},11}/\sigma^{3}=\SI{33,10}{}$, and $\Omega_{\mathrm{S},22}/\sigma^{3}=\SI{604,96}{}$. Here, the agreement is good. The maximal deviation from the simulation results is below $\SI{0.4}{\%}$ for elements $K_{ii}$ and $\SI{5}{\%}$ for elements $\Omega_{\mathrm{S},ii}$ with $i\in\{1,2\}$.

\subsection{\label{sec:cylindersphericalconical}Right circular cylinder with concave and convex spherical or conical ends}
\begin{figure*}[htb]
\centering
\includegraphics[width=0.75\linewidth]{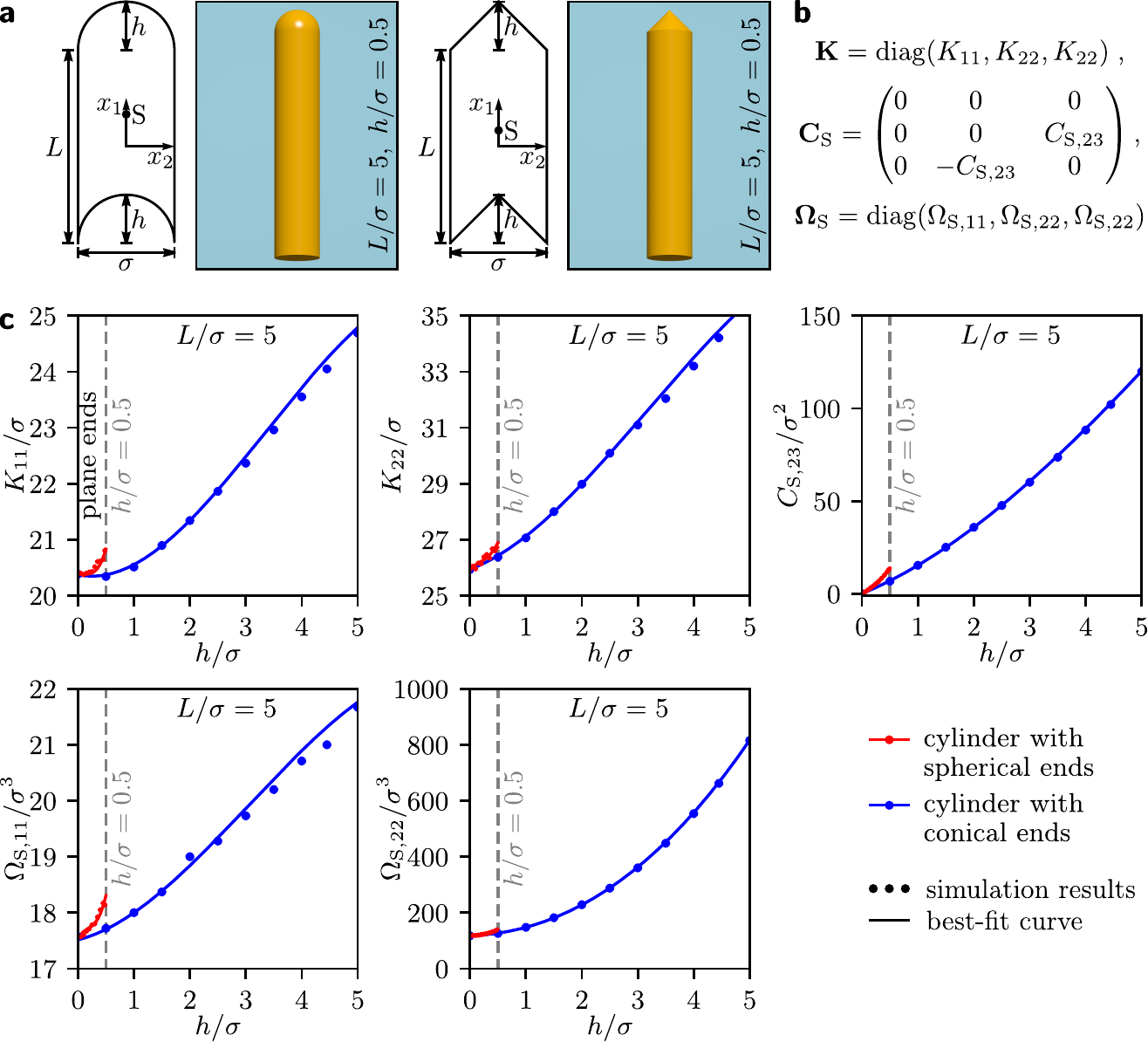}%
\caption{\label{fig:cylinder_spherical_conical}Analogous to Fig.\ \ref{fig:cylinder_cuboid}, but now for a right circular cylinder with concave and convex spherical or conical ends, diameter $\sigma$, length $L$ of the cylindrical part, and cap height $h$.
Here, the nonvanishing elements of $\mathbf{K}$, $\mathbf{C}_{\mathrm{S}}$, and $\mathbf{\Omega}_{\mathrm{S}}$ are functions of both $L/\sigma$ and $h/\sigma$ and shown for $L/\sigma=5$; the corresponding best-fit curves are given by $p_{2}(L/\sigma,h/\sigma)$ with the coefficients from Tabs.\ \ref{tab:cylinder_spherical} (for spherical ends) and \ref{tab:cylinder_conical} (for conical ends). 
The data for the limiting case $h=0$ belong to the cylinder with plane ends from Fig.\ \ref{fig:cylinder_cuboid}; For geometric reasons, the data for the cylinder with concave and convex spherical ends stop at $h/\sigma=0.5$.}
\end{figure*}

The next particle shapes we consider are right circular cylinders with concave and convex spherical or conical ends (see Fig.\ \ref{fig:cylinder_spherical_conical}).
Both shapes have the same symmetry properties. They have rotational symmetry about the $x_{1}$ axis, but no reflection symmetry with respect to a plane perpendicular to their symmetry axis. Therefore, their shape is \ZT{polar}. Through these symmetry properties, the hydrodynamic resistance matrix $\HH$ has five independent nonzero elements. These are the elements $K_{11}$, $K_{22}$, $C_{\mathrm{S},23}$, $\Omega_{\mathrm{S},11}$, and $\Omega_{\mathrm{S},22}$. Due to the broken reflection symmetry with respect to the $x_{2}$-$x_{3}$ plane, the shapes have a translational-rotational coupling described by the element $C_{\mathrm{S},23}$. 
We calculated the nonzero elements of the hydrodynamic resistance matrices for the two particle shapes as functions of $L/\sigma\in [0.5,10]$ and $h/\sigma\in [0,5]$, where $L$ is the length of the cylindrical part, $\sigma$ is the diameter, and $h$ is the height of the end caps of the particle. 
In Fig.\ \ref{fig:cylinder_spherical_conical}, the direct simulation results and corresponding best-fit curves are shown for fixed $L/\sigma=5$ and variable $h/\sigma$. 
The two-dimensional best-fit curves are given by the polynomial \eqref{eq:pII} with the best-fit values for the coefficients of the polynomial that are given in Tabs.\ \ref{tab:cylinder_spherical} and \ref{tab:cylinder_conical} for the spherical and conical ends, respectively.
Again, the agreement of the direct simulation results and best-fit curves is very good. 
For given values of the length $L$ and diameter $\sigma$, all nonzero elements of $\HH$ increase with $h$. This applies to both particle shapes and can be understood from the fact that with $h$ the surface of the particle and thus the hydrodynamic drag grow.
The data for the limiting case $h=0$, where all nonzero elements of $\HH$ attain their smallest values, belong to the cylinder with plane ends from Fig.\ \ref{fig:cylinder_cuboid}. 
At $h=0$, the curves for the cylinder with spherical ends have thus the same values as the curves for the cylinder with conical ends. For $h>0$, the former curves have larger values and are faster growing than the latter curves. This is in line with the fact that for a given diameter $\sigma$ and nonzero height $h$, a spherical end has a by a factor $\sqrt{1+(2h/\sigma)^{2}}$ larger surface than a conical end. Since the spherical ends cannot be higher than a half sphere, the data for the cylinder with concave and convex spherical ends stop at $h/\sigma=0.5$.

\subsection{\label{sec:halfballandcone}Half ball and right circular cone}
\begin{figure*}[htb]
\centering
\includegraphics[width=0.75\linewidth]{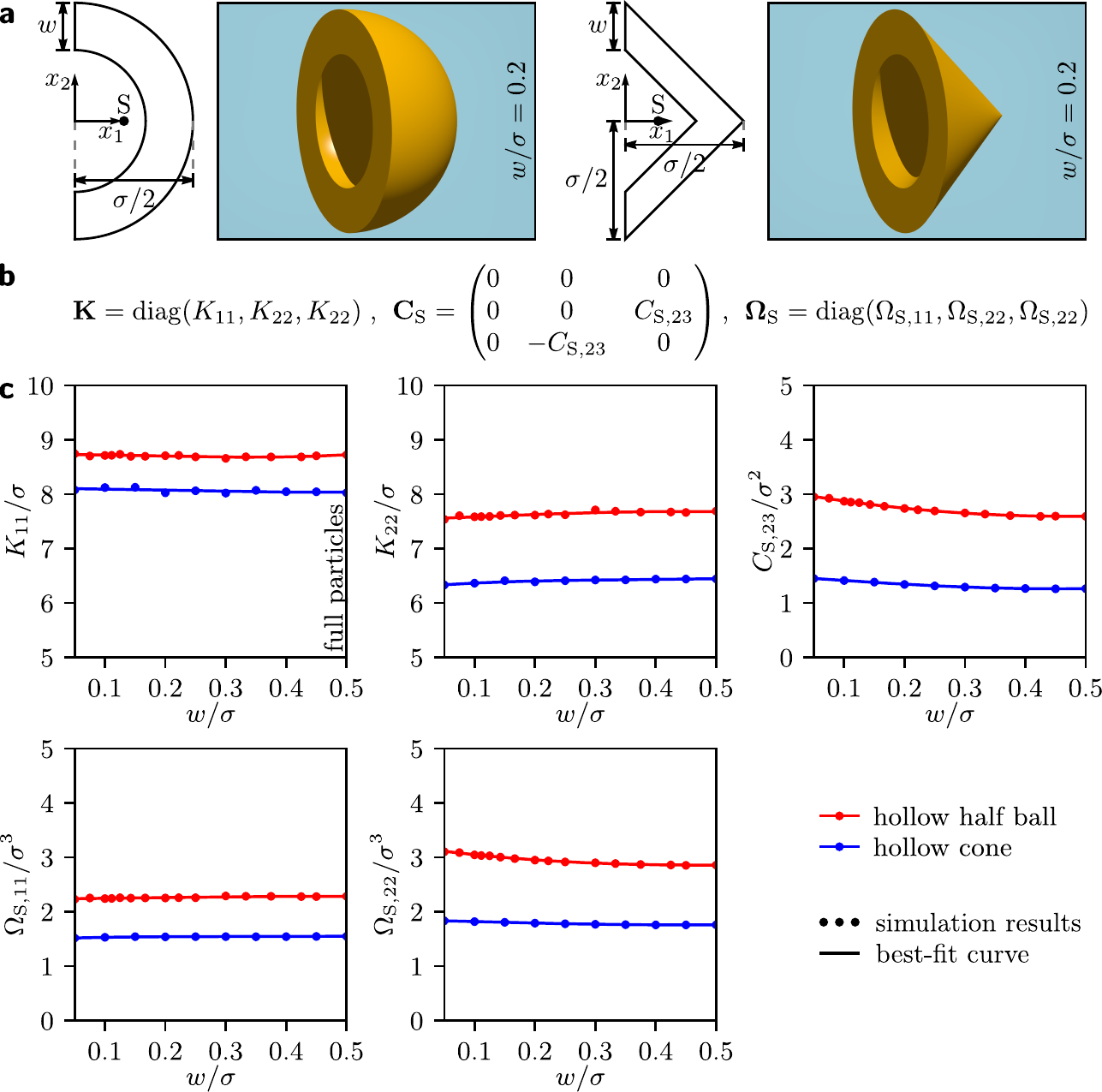}%
\caption{\label{fig:half_ball_cone}Analogous to Fig.\ \ref{fig:cylinder_cuboid}, but now for a hollow half ball and a hollow right circular cone with diameter $\sigma$, height $\sigma/2$, and wall thickness $w$. Here, the nonvanishing elements of $\mathbf{K}$, $\mathbf{C}_{\mathrm{S}}$, and $\mathbf{\Omega}_{\mathrm{S}}$ are functions of $w/\sigma$; the corresponding best-fit curves are given by $p_{1}(w/\sigma)$ with the coefficients from Tabs.\ \ref{tab:half_ball} (for half ball) and \ref{tab:cone} (for cone). The data for the limiting case $w/\sigma=0.5$ belong to a full half ball and a full right circular cone.}
\end{figure*}

We continue with results for the hydrodynamic resistance matrices of half balls and right circular cones that are hollow or full (see Fig.\ \ref{fig:half_ball_cone}).
Both the half balls and right circular cones have the same symmetry properties as the polar cylinders from section \ref{sec:cylindersphericalconical}.
The shapes considered here are thus polar as well. As a further consequence of the equivalence of the symmetry properties, the hydrodynamic resistance matrix $\HH$ has the same structure as in the previous section.
We studied how the nonzero elements of the hydrodynamic resistance matrices for the hollow half balls and right circular cones depend on $w/\sigma\in [0.05,0.5]$, where $\sigma$ is the diameter and $w$ is the wall thickness of a particle. For equal height-to-diameter ratios of half balls and cones, we set the height of the cones to $\sigma/2$.  
The best-fit curves are now given by the polynomial \eqref{eq:pI} with the best-fit values for the coefficients of the polynomial that are given in Tabs.\ \ref{tab:half_ball} and \ref{tab:cone} for the half balls and cones, respectively.
As for the particle shapes studied in previous sections, there is a very good agreement of the direct simulation results and corresponding best-fit curves. 
The course of the curves is rather simple.
For a given $\sigma$, all elements of $\HH$ are nearly independent of $w$. This means that the hollow particles have nearly the same hydrodynamic resistance matrices as the corresponding full particles, which are obtained in the limiting case $w/\sigma=0.5$. The values of the nonzero elements of $\HH$ are always larger for the half balls than for the cones. 
This is consistent with the observation from section \ref{sec:cylindersphericalconical} that the nonzero elements are larger for a right circular cylinder with concave and convex spherical ends than for one with conical ends.

\subsection{Double-cup shapes}
\begin{figure*}[htb]
\centering
\includegraphics[width=\linewidth]{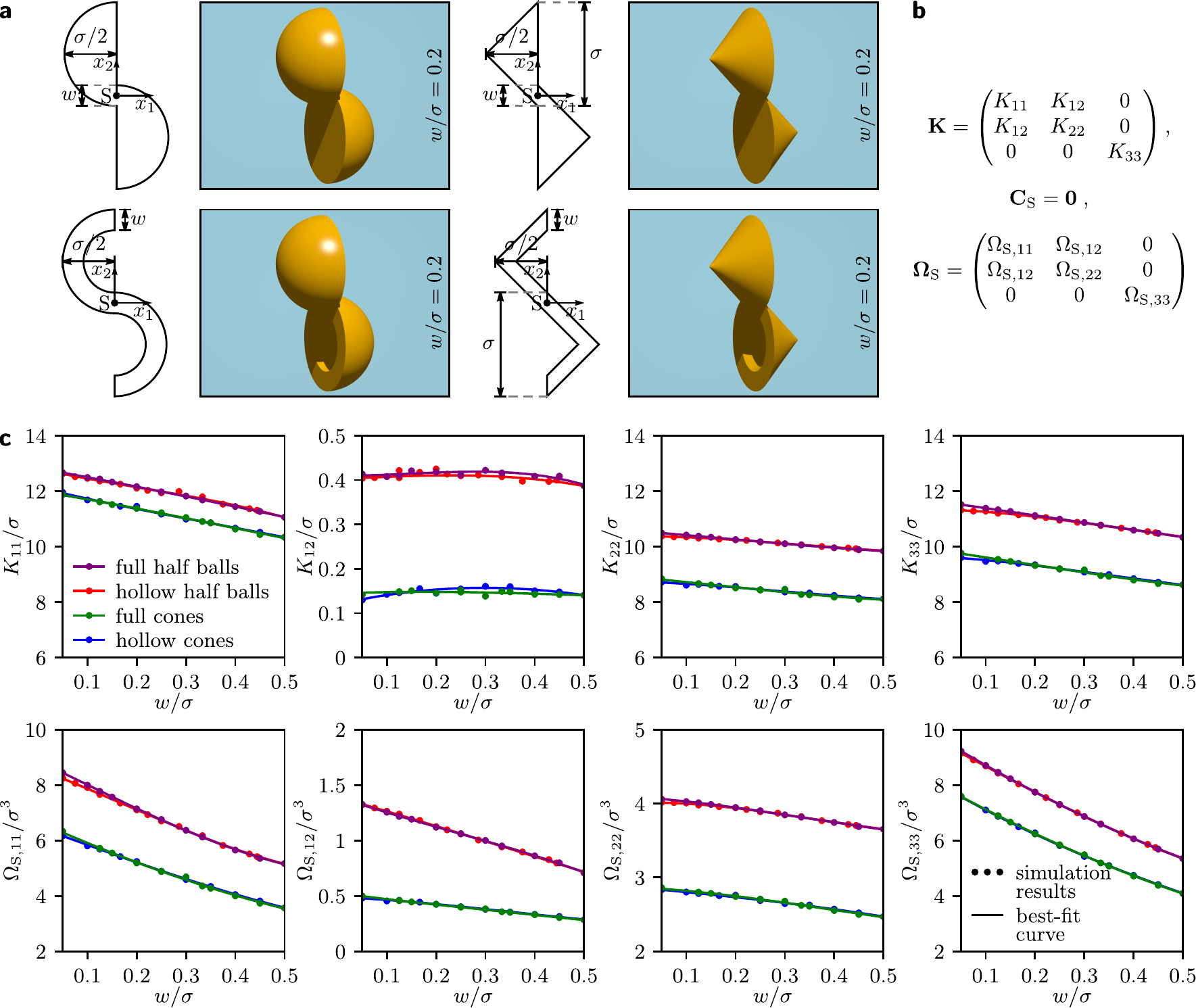}%
\caption{\label{fig:double_cups}Analogous to Fig.\ \ref{fig:half_ball_cone}, but now for double-cup particles consisting of two of the half balls or right circular cones from Fig.\ \ref{fig:half_ball_cone}. In the case of hollow constituent particles, the width $w$ of their contact area equals their wall thickness. Here, the nonvanishing elements of $\mathbf{K}$, $\mathbf{C}_{\mathrm{S}}$, and $\mathbf{\Omega}_{\mathrm{S}}$ are functions of $w/\sigma$; The corresponding best-fit curves are given by $p_{1}(w/\sigma)$ with the coefficients from Tabs.\ \ref{tab:double_cup_half_balls} (for full half balls), \ref{tab:double_cup_cones} (for full cones), \ref{tab:double_cup_half_balls_hollow} (for hollow half balls), and \ref{tab:double_cup_cones_hollow} (for hollow cones).}
\end{figure*}

Finally, we consider four qualitatively different double-cup particles, each of them consisting of two of the hollow or full half balls or right circular cones from section \ref{sec:halfballandcone} (see Fig.\ \ref{fig:double_cups}). 
All these double-cup particles have the same symmetry properties. That are a reflection symmetry with respect to the $x_{1}$-$x_{2}$ plane and a two-fold rotational symmetry with respect to the $x_{3}$ axis. The associated symmetry group is therefore $\mathrm{C_{2h}}$. 
Owing to these symmetry properties, the hydrodynamic resistance matrix $\HH$ has eight independent nonzero elements for the double-cup particles. 
These elements are $K_{11}$, $K_{12}$, $K_{22}$, $K_{33}$, $\Omega_{\mathrm{S},11}$, $\Omega_{\mathrm{S},12}$, $\Omega_{\mathrm{S},22}$, and $\Omega_{\mathrm{S},33}$. 
Although there is no coupling of translational and rotational motion, for both types of motion there is a coupling between the directions $x_{1}$ and $x_{2}$. This coupling is described by the elements $K_{12}$ for translational and $\Omega_{\mathrm{S},12}$ for rotational motion. 
Figure \ref{fig:double_cups} shows the obtained direct simulation results and corresponding best-fit curves for the nonzero elements of the hydrodynamic resistance matrices of the double-cup particles as functions of $w/\sigma\in [0.05,0.5]$, where $w$ is the width of the contact area of the two constituent particles. 
When the constituent particles are hollow, $w$ is also equal to the thickness of their walls. 
The best-fit curves are given by the polynomial \eqref{eq:pI} with the best-fit values for the coefficients of the polynomial that are given in Tabs.\ \ref{tab:double_cup_half_balls}-\ref{tab:double_cup_cones_hollow} for the full half balls, full cones, hollow half balls, and hollow cones as constituent particles, respectively.
Also for the double-cup shapes, a very good agreement of the direct simulation results and best-fit curves is visible. 
The curves are rather straight with only small curvatures. When $\sigma$ is kept constant and $w$ is increased, $K_{12}$ stays nearly unchanged whereas the other nonzero elements of $\HH$ decrease. This behavior can be observed for all four double-cup shapes. 
The decrease is in line with the fact that for growing $w$ the size and surface of the double-cup particles and thus the hydrodynamic drag on them decline.
Corresponding curves for hollow and full constituent particles are so close that they are difficult to distinguish. This is consistent with related findings for the particles in section \ref{sec:halfballandcone}. 
In the limiting case $w/\sigma=0.5$, the hollow particles become full and the values of the associated curves become equal. The curves for half balls and cones as constituents look similar, but the curves for the half balls are always clearly above the curves for the cones. Also this is consistent with analogous findings in section \ref{sec:halfballandcone}.

\section{\label{conclusions}Conclusions}
Based on numerically solving the Stokes equation for low-Reynolds-number flows with the finite element method, we have calculated the hydrodynamic resistance matrices of colloidal particles with various shapes. Since a hydrodynamic resistance matrix describes the shape- and size-dependent hydrodynamic properties of a free or driven rigid particle in a simple liquid at low Reynolds number, this matrix is frequently needed when addressing the Brownian or deterministic dynamics of such a particle by theoretical methods like, e.g., Langevin equations, Smoluchowski equations, and classical dynamical density functional theory for colloidal particles with anisometric shapes. 
The considered shapes include apolar and polar as well as convex and partially concave ones. They range from apolar rodlike shapes, which are a standard choice when studying nonspherical particles, to more complex shapes with particular symmetry properties that have gained increasing attention in recent years \cite{ZerroukiBPCB2008,SokolovAGA2010,WangCHM2012,GarciaGradillaEtAl2013,MijalkovV2013,SchamelPGMMF2013,KuemmeltHWBEVLB2013,KuemmeltHWTBEVLB2014,tenHagenKWTLB2014,YuanMSTS2018}.  
The presented results are quite accurate and well in line with available analytical and numerical comparative data. 
We therefore believe that this work will stimulate and support a lot of future studies that focus on the dynamics of anisometric colloidal particles. Among the expectable future applications of our results are Brownian dynamics simulations of passive and active colloidal liquid crystals, where the hydrodynamic resistance matrix of the colloidal particles is needed to simulate their Brownian motion correctly.

\section*{Conflicts of interest}
There are no conflicts of interest to declare.

\begin{acknowledgments}
R.W.\ is funded by the Deutsche Forschungsgemeinschaft (DFG, German Research Foundation) -- WI 4170/3-1. 
The simulations for this work were performed on the computer cluster PALMA of the University of M\"unster. 
\end{acknowledgments}

\FloatBarrier
\onecolumngrid
\appendix
\section{\label{appendix}Best-fit coefficients for the considered particles}
The best-fit values of the coefficients of the functions \eqref{eq:pI} and \eqref{eq:pII}, which describe the nonzero elements of the hydrodynamic resistance matrix \eqref{eq:H} as functions of one or two shape parameters, respectively, are listed in the following tables for all considered particle shapes. In addition, the root-mean-square deviation of the underlying fit is given for each set of coefficient values. The data correspond to particles with their centers of mass as reference points and with their orientations as shown in Figs.\ \ref{fig:cylinder_cuboid}-\ref{fig:double_cups}. For an easier use of the data, they are also available as a supplementary spreadsheet file. 

\begin{table*}[htbp]
\centering
\begin{tabular}{|c|c|c|c|c|}
\hline
& $K_{11}$ & $K_{22}$ & $\Omega_{\mathrm{S},11}$ & $\Omega_{\mathrm{S},22}$ \\\hline 
$a_{0}$& \SI{8.30E+00}{}& \SI{6.41E+00}{}& \SI{1.71E+00}{}& \SI{2.96E+00}{}\\\hline 
$a_{1}$& \SI{2.97E+00}{}& \SI{4.88E+00}{}& \SI{3.00E+00}{}& \SI{-8.71E-01}{}\\\hline 
$a_{2}$& \SI{-1.38E-01}{}& \SI{-2.43E-01}{}& \SI{5.42E-02}{}& \SI{3.40E+00}{}\\\hline 
$a_{3}$& \SI{5.23E-03}{}& \SI{9.84E-03}{}& \SI{-4.00E-03}{}& \SI{2.71E-01}{}\\\hline 
$\mathrm{RMSD}$& \SI{4.50E-02}{}& \SI{9.48E-02}{}& \SI{1.31E-01}{}& \SI{4.98E-01}{}\\\hline
\end{tabular} 
\caption{\label{tab:cylinder}Best-fit coefficients of the function $p_{1}(L/\sigma)$ with fit range $L/\sigma\in [0.5,10]$ and the corresponding root-mean-square deviation (RMSD) for each nonzero element of the hydrodynamic resistance matrix $\HH$ of a \textit{right circular cylinder with plane ends}, which has diameter $\sigma$ and length $L$ (see Fig.\ \ref{fig:cylinder_cuboid}a). Depending on the aspect ratio $L/\sigma$, the results were obtained using a mesh with \SI{1.46E+04} to \SI{2.07E+05} triangles representing the particle surface and \SI{2.44E+06} to \SI{3.02E+06} tetrahedra in the domain outside of the particle.}
\end{table*}
 
\begin{table*}[htbp]
\centering
\begin{tabular}{|c|c|c|c|c|}
\hline
& $K_{11}$ & $K_{22}$ & $\Omega_{\mathrm{S},11}$ & $\Omega_{\mathrm{S},22}$ \\\hline 
$a_{0}$& \SI{9.49E+00}{}& \SI{7.43E+00}{}& \SI{2.70E+00}{}& \SI{3.04E+00}{}\\\hline 
$a_{1}$& \SI{2.98E+00}{}& \SI{5.09E+00}{}& \SI{5.03E+00}{}& \SI{7.49E-01}{}\\\hline 
$a_{2}$& \SI{-1.38E-01}{}& \SI{-2.55E-01}{}& \SI{-3.45E-02}{}& \SI{3.51E+00}{}\\\hline 
$a_{3}$& \SI{5.32E-03}{}& \SI{1.03E-02}{}& \SI{1.67E-03}{}& \SI{3.04E-01}{}\\\hline 
$\mathrm{RMSD}$& \SI{6.30E-02}{}& \SI{1.04E-01}{}& \SI{6.17E-02}{}& \SI{1.71E-02}{}\\\hline 
\end{tabular} 
\caption{\label{tab:cuboid}Analogous to Tab.\ \ref{tab:cylinder}, but now for a \textit{rectangular cuboid with quadratic cross section}, which has width $\sigma$ and length $L$ (see Fig.\ \ref{fig:cylinder_cuboid}a). The used mesh included \SI{6.98E+03} to \SI{8.03E+04} triangles on the particle surface and \SI{3.26E+06} to \SI{3.56E+06} tetrahedra outside of the particle.}
\end{table*}

\begin{table*}[htbp]
\centering
\begin{tabular}{|c|c|c|c|c|c|}
\hline
& $K_{11}$ & $K_{22}$ & $C_{\mathrm{S},23}$ & $\Omega_{\mathrm{S},11}$ & $\Omega_{\mathrm{S},22}$ \\\hline 
$a_{0}$& \SI{8.43E+00}{}& \SI{6.53E+00}{}& \SI{-1.81E-01}{}& \SI{1.55E+00}{}& \SI{3.83E+00}{}\\\hline 
$a_{1}$& \SI{2.93E+00}{}& \SI{4.84E+00}{}& \SI{1.48E-01}{}& \SI{3.27E+00}{}& \SI{-9.64E-01}{}\\\hline 
$a_{2}$& \SI{3.90E-01}{}& \SI{1.93E+00}{}& \SI{6.54E+00}{}& \SI{1.66E+00}{}& \SI{1.36E+01}{}\\\hline 
$a_{3}$& \SI{-1.34E-01}{}& \SI{-2.36E-01}{}& \SI{-3.59E-02}{}& \SI{-1.81E-02}{}& \SI{3.36E+00}{}\\\hline 
$a_{4}$& \SI{8.50E-02}{}& \SI{-2.93E-01}{}& \SI{3.01E+00}{}& \SI{-2.06E-02}{}& \SI{-1.72E+00}{}\\\hline 
$a_{5}$& \SI{-3.72E+00}{}& \SI{1.05E-01}{}& \SI{-9.39E+00}{}& \SI{-5.24E+00}{}& \SI{-7.87E+01}{}\\\hline 
$a_{6}$& \SI{5.13E-03}{}& \SI{9.34E-03}{}& \SI{2.44E-03}{}& \SI{8.00E-04}{}& \SI{2.75E-01}{}\\\hline 
$a_{7}$& \SI{-1.15E-02}{}& \SI{2.33E-02}{}& \SI{-8.24E-02}{}& \SI{-2.39E-03}{}& \SI{4.32E-01}{}\\\hline 
$a_{8}$& \SI{1.83E-01}{}& \SI{-1.13E-01}{}& \SI{2.40E+00}{}& \SI{8.82E-02}{}& \SI{1.61E+01}{}\\\hline 
$a_{9}$& \SI{7.40E+00}{}& \SI{4.05E+00}{}& \SI{2.74E+01}{}& \SI{9.70E+00}{}& \SI{1.16E+02}{}\\\hline 
$\mathrm{RMSD}$& \SI{1.62E-01}{}& \SI{1.08E-01}{}& \SI{9.88E-02}{}& \SI{1.46E-01}{}& \SI{1.31E+00}{}\\\hline 
\end{tabular} 
\caption{\label{tab:cylinder_spherical}Best-fit coefficients of the function $p_{2}(L/\sigma,h/\sigma)$ with fit ranges $L/\sigma\in [0.5,10]$ and $h/\sigma\in [0,5]$ as well as the corresponding root-mean-square deviation (RMSD) for each nonzero element of the hydrodynamic resistance matrix $\HH$ of a \textit{right circular cylinder with concave and convex spherical ends}, which has diameter $\sigma$, length of the cylindrical part $L$, and cap height $h$ (see Fig.\ \ref{fig:cylinder_spherical_conical}a). Depending on $L/\sigma$ and $h/\sigma$, the results were obtained using a mesh with \SI{1.46E+04} to \SI{2.14E+05} triangles representing the particle surface and \SI{2.44E+06} to \SI{3.66E+06} tetrahedra in the domain outside of the particle.}
\end{table*}
 
\begin{table*}[htbp]
\centering
\begin{tabular}{|c|c|c|c|c|c|}
\hline
& $K_{11}$ & $K_{22}$ & $C_{\mathrm{S},23}$ & $\Omega_{\mathrm{S},11}$ & $\Omega_{\mathrm{S},22}$ \\\hline 
$a_{0}$& \SI{8.36E+00}{}& \SI{6.82E+00}{}& \SI{-6.68E-02}{}& \SI{1.48E+00}{}& \SI{4.13E+00}{}\\\hline 
$a_{1}$& \SI{2.95E+00}{}& \SI{4.64E+00}{}& \SI{8.94E-02}{}& \SI{3.35E+00}{}& \SI{-8.59E-01}{}\\\hline 
$a_{2}$& \SI{1.74E-01}{}& \SI{1.45E+00}{}& \SI{2.20E+00}{}& \SI{3.70E-01}{}& \SI{-8.69E+00}{}\\\hline 
$a_{3}$& \SI{-1.34E-01}{}& \SI{-1.99E-01}{}& \SI{-1.67E-02}{}& \SI{-4.11E-02}{}& \SI{3.35E+00}{}\\\hline 
$a_{4}$& \SI{-9.41E-02}{}& \SI{-2.07E-01}{}& \SI{2.24E+00}{}& \SI{-5.06E-03}{}& \SI{1.42E+00}{}\\\hline 
$a_{5}$& \SI{3.78E-01}{}& \SI{4.80E-01}{}& \SI{2.84E+00}{}& \SI{2.07E-01}{}& \SI{8.22E+00}{}\\\hline 
$a_{6}$& \SI{5.01E-03}{}& \SI{7.37E-03}{}& \SI{5.56E-04}{}& \SI{2.50E-03}{}& \SI{2.72E-01}{}\\\hline 
$a_{7}$& \SI{5.13E-03}{}& \SI{1.20E-02}{}& \SI{-3.27E-02}{}& \SI{-7.34E-04}{}& \SI{4.08E-01}{}\\\hline 
$a_{8}$& \SI{5.02E-03}{}& \SI{6.41E-03}{}& \SI{3.21E-03}{}& \SI{2.04E-03}{}& \SI{2.56E+00}{}\\\hline 
$a_{9}$& \SI{-3.87E-02}{}& \SI{-5.42E-02}{}& \SI{-1.14E-01}{}& \SI{-2.27E-02}{}& \SI{1.04E+00}{}\\\hline 
$\mathrm{RMSD}$& \SI{7.28E-03}{}& \SI{4.64E-02}{}& \SI{4.95E-02}{}& \SI{2.48E-02}{}& \SI{4.28E+00}{}\\\hline 
\end{tabular} 
\caption{\label{tab:cylinder_conical}Analogous to Tab.\ \ref{tab:cylinder_spherical}, but now for a \textit{right circular cylinder with concave and convex conical ends}, which has diameter $\sigma$, length of the cylindrical part $L$, and cap height $h$ (see Fig.\ \ref{fig:cylinder_spherical_conical}a). The used mesh included \SI{1.46E+04} to \SI{2.80E+05} triangles on the particle surface and \SI{2.44E+06} to \SI{3.24E+06} tetrahedra outside of the particle.}
\end{table*}

\begin{table*}[htbp]
\centering
\begin{tabular}{|c|c|c|c|c|c|}
\hline
& $K_{11}$ & $K_{22}$ & $C_{\mathrm{S},23}$ & $\Omega_{\mathrm{S},11}$ & $\Omega_{\mathrm{S},22}$ \\\hline 
$a_{0}$& \SI{8.73E+00}{}& \SI{7.53E+00}{}& \SI{3.05E+00}{}& \SI{2.23E+00}{}& \SI{3.18E+00}{}\\\hline 
$a_{1}$& \SI{-2.25E-02}{}& \SI{5.80E-01}{}& \SI{-2.02E+00}{}& \SI{6.31E-02}{}& \SI{-1.59E+00}{}\\\hline 
$a_{2}$& \SI{-1.21E+00}{}& \SI{-2.12E-01}{}& \SI{2.38E+00}{}& \SI{5.27E-01}{}& \SI{2.39E+00}{}\\\hline 
$a_{3}$& \SI{2.46E+00}{}& \SI{-7.16E-01}{}& \SI{-3.60E-01}{}& \SI{-9.27E-01}{}& \SI{-1.04E+00}{}\\\hline 
$\mathrm{RMSD}$& \SI{1.28E-02}{}& \SI{1.88E-02}{}& \SI{4.84E-03}{}& \SI{7.26E-03}{}& \SI{4.64E-03}{}\\\hline
\end{tabular} 
\caption{\label{tab:half_ball}Best-fit coefficients of the function $p_{1}(w/\sigma)$ with fit range $w/\sigma\in [0.05,0.5]$ and the corresponding root-mean-square deviation (RMSD) for each nonzero element of the hydrodynamic resistance matrix $\HH$ of a \textit{hollow half ball}, which has diameter $\sigma$ and wall thickness $w$ (see Fig.\ \ref{fig:half_ball_cone}a) and becomes a \textit{full half ball} in the limiting case $w/\sigma=0.5$. Depending on $w/\sigma$, the results were obtained using a mesh with \SI{6.54E+04} to \SI{8.21E+04} triangles representing the particle surface and \SI{2.25E+06} to \SI{2.81E+06} tetrahedra in the domain outside of the particle.}
\end{table*}

\begin{table*}[htbp]
\centering
\begin{tabular}{|c|c|c|c|c|c|}
\hline
& $K_{11}$ & $K_{22}$ & $C_{\mathrm{S},23}$ & $\Omega_{\mathrm{S},11}$ & $\Omega_{\mathrm{S},22}$ \\\hline 
$a_{0}$& \SI{8.10E+00}{}& \SI{6.29E+00}{}& \SI{1.49E+00}{}& \SI{1.50E+00}{}& \SI{1.85E+00}{}\\\hline 
$a_{1}$& \SI{-4.72E-02}{}& \SI{9.08E-01}{}& \SI{-7.85E-01}{}& \SI{4.39E-01}{}& \SI{-3.01E-01}{}\\\hline
$a_{2}$& \SI{-8.10E-01}{}& \SI{-2.23E+00}{}& \SI{6.15E-02}{}& \SI{-1.45E+00}{}& \SI{-5.61E-02}{}\\\hline 
$a_{3}$& \SI{1.26E+00}{}& \SI{2.03E+00}{}& \SI{1.21E+00}{}& \SI{1.55E+00}{}& \SI{6.12E-01}{}\\\hline 
$\mathrm{RMSD}$& \SI{2.76E-02}{}& \SI{9.08E-03}{}& \SI{2.66E-03}{}& \SI{1.79E-03}{}& \SI{6.57E-04}{}\\\hline
\end{tabular} 
\caption{\label{tab:cone}Analogous to Tab.\ \ref{tab:half_ball}, but now for a \textit{hollow right circular cone}, which has diameter $\sigma$ and wall thickness $w$ (see Fig.\ \ref{fig:half_ball_cone}a) and becomes a \textit{full right circular cone} in the limiting case $w/\sigma=0.5$. The used mesh included \SI{5.24E+04} to \SI{5.91E+04} triangles on the particle surface and \SI{2.29E+06} to \SI{2.40E+06} tetrahedra outside of the particle.}
\end{table*}
            
\begin{table*}[htbp]
\centering
\begin{tabular}{|c|c|c|c|c|c|c|c|c|}
\hline
& $K_{11}$ &  $K_{12}$ & $K_{22}$ & $K_{33}$ & $\Omega_{\mathrm{S},11}$ & $\Omega_{\mathrm{S},12}$ & $\Omega_{\mathrm{S},22}$ & $\Omega_{\mathrm{S},33}$ \\\hline 
$a_{0}$& \SI{1.28E+01}{}& \SI{4.11E-01}{}& \SI{1.06E+01}{}& \SI{1.16E+01}{} & \SI{8.88E+00}{} & \SI{1.39E+00}{} & \SI{4.08E+00}{}& \SI{9.75E+00}{}\\\hline 
$a_{1}$& \SI{-3.24E+00}{}& \SI{-2.67E-02}{}& \SI{-1.37E+00}{}& \SI{-2.41E+00}{} & \SI{-8.67E+00}{} & \SI{-1.50E+00}{} & \SI{-4.00E-01}{}& \SI{-1.08E+01}{}\\\hline 
$a_{2}$& \SI{-4.26E-01}{}& \SI{4.81E-01}{}& \SI{-1.18E+00}{}& \SI{-8.25E-01}{} & \SI{-1.46E+00}{} & \SI{1.06E+00}{} & \SI{-1.87E+00}{}& \SI{3.67E+00}{}\\\hline 
$a_{3}$& \SI{-3.02E-01}{}& \SI{-1.02E+00}{}& \SI{2.06E+00}{}& \SI{9.30E-01}{} & \SI{7.73E+00}{} & \SI{-1.59E+00}{} & \SI{1.84E+00}{}& \SI{4.87E-01}{}\\\hline 
$\mathrm{RMSD}$& \SI{8.93E-03}{}& \SI{4.03E-03}{}& \SI{3.58E-03}{}& \SI{5.84E-03}{} & \SI{1.55E-02}{}& \SI{4.02E-03}{}& \SI{3.42E-03}{}& \SI{3.60E-03}{}\\\hline
\end{tabular} 
\caption{\label{tab:double_cup_half_balls}Analogous to Tab.\ \ref{tab:half_ball}, but now for a \textit{double-cup particle consisting of two half balls}, which have diameter $\sigma$ and width of the contact area $w$ (see Fig.\ \ref{fig:double_cups}a). The used mesh included \SI{1.14E+05} to \SI{1.30E+05} triangles on the particle surface and \SI{2.59E+06} to \SI{2.80E+06} tetrahedra outside of the particle.}
\end{table*}

\begin{table*}[htbp]
\centering
\begin{tabular}{|c|c|c|c|c|c|c|c|c|}
\hline
& $K_{11}$ &  $K_{12}$ & $K_{22}$ & $K_{33}$ & $\Omega_{\mathrm{S},11}$ & $\Omega_{\mathrm{S},12}$ & $\Omega_{\mathrm{S},22}$ & $\Omega_{\mathrm{S},33}$ \\\hline 
$a_{0}$& \SI{1.20E+01}{}& \SI{1.44E-01}{}& \SI{8.90E+00}{}& \SI{9.88E+00}{}& \SI{6.67E+00}{}& \SI{5.26E-01}{}& \SI{2.87E+00}{}& \SI{8.06E+00}{}\\\hline  
$a_{1}$& \SI{-2.58E+00}{}& \SI{4.35E-02}{}& \SI{-1.79E+00}{}& \SI{-2.58E+00}{}& \SI{-7.90E+00}{}& \SI{-5.73E-01}{}& \SI{-2.91E-01}{}& \SI{-9.74E+00}{}\\\hline 
$a_{2}$& \SI{-3.11E+00}{}& \SI{-1.46E-01}{}& \SI{-7.35E-01}{}& \SI{-1.14E+00}{}& \SI{2.78E+00}{}& \SI{3.52E-01}{}& \SI{-1.79E+00}{}& \SI{3.30E+00}{}\\\hline 
$a_{3}$& \SI{3.04E+00}{}& \SI{8.71E-02}{}& \SI{2.09E+00}{}& \SI{2.34E+00}{}& \SI{1.01E+00}{}& \SI{-3.61E-01}{}& \SI{1.52E+00}{}& \SI{5.45E-01}{}\\\hline 
$\mathrm{RMSD}$& \SI{2.27E-02}{}& \SI{3.34E-03}{}& \SI{3.55E-02}{}& \SI{3.14E-02}{}& \SI{3.59E-02}{}& \SI{2.48E-03}{}& \SI{8.82E-03}{}& \SI{1.39E-02}{}\\\hline 
\end{tabular} 
\caption{\label{tab:double_cup_cones}Analogous to Tab.\ \ref{tab:half_ball}, but now for a \textit{double-cup particle consisting of two right circular cones}, which have diameter $\sigma$ and width of the contact area $w$ (see Fig.\ \ref{fig:double_cups}a). The used mesh included \SI{8.77E+04} to \SI{1.03E+05} triangles on the particle surface and \SI{2.31E+06} to \SI{2.51E+06} tetrahedra outside of the particle.}
\end{table*}

\begin{table*}[htbp]
\centering
\begin{tabular}{|c|c|c|c|c|c|c|c|c|}
\hline
& $K_{11}$ &  $K_{12}$ & $K_{22}$ & $K_{33}$ & $\Omega_{\mathrm{S},11}$ & $\Omega_{\mathrm{S},12}$ & $\Omega_{\mathrm{S},22}$ & $\Omega_{\mathrm{S},33}$ \\\hline 
$a_{0}$& \SI{1.28E+01}{}& \SI{3.97E-01}{}& \SI{1.04E+01}{}& \SI{1.13E+01}{}& \SI{8.57E+00}{}& \SI{1.39E+00}{}& \SI{4.01E+00}{}& \SI{9.63E+00}{}\\\hline 
$a_{1}$& \SI{-4.05E+00}{}& \SI{2.17E-01}{}& \SI{-5.73E-02}{}& \SI{-2.84E-01}{}& \SI{-6.52E+00}{}& \SI{-1.26E+00}{}& \SI{2.41E-01}{}& \SI{-9.71E+00}{}\\\hline  
$a_{2}$& \SI{5.31E+00}{}& \SI{-7.41E-01}{}& \SI{-3.86E+00}{}& \SI{-6.13E+00}{}& \SI{-6.12E+00}{}& \SI{-2.69E-01}{}& \SI{-3.58E+00}{}& \SI{6.91E-01}{}\\\hline 
$a_{3}$& \SI{-8.51E+00}{}& \SI{5.45E-01}{}& \SI{3.74E+00}{}& \SI{5.47E+00}{}& \SI{1.10E+01}{}& \SI{1.75E-01}{}& \SI{3.36E+00}{}& \SI{3.23E+00}{}\\\hline 
$\mathrm{RMSD}$& \SI{4.10E-02}{}& \SI{4.79E-03}{}& \SI{8.12E-03}{}& \SI{1.35E-02}{}& \SI{2.32E-02}{}& \SI{6.06E-03}{}& \SI{5.66E-03}{}& \SI{9.03E-03}{}\\\hline 
\end{tabular} 
\caption{\label{tab:double_cup_half_balls_hollow}Analogous to Tab.\ \ref{tab:half_ball}, but now for a \textit{double-cup particle consisting of two hollow half balls}, which have diameter $\sigma$ and wall thickness $w$ (see Fig.\ \ref{fig:double_cups}a). The used mesh included \SI{1.14E+05} to \SI{1.65E+05} triangles on the particle surface and \SI{2.59E+06} to \SI{3.99E+06} tetrahedra outside of the particle.}
\end{table*}

\begin{table*}[htbp]
\centering
\begin{tabular}{|c|c|c|c|c|c|c|c|c|}
\hline
& $K_{11}$ &  $K_{12}$ & $K_{22}$ & $K_{33}$ & $\Omega_{\mathrm{S},11}$ & $\Omega_{\mathrm{S},12}$  & $\Omega_{\mathrm{S},22}$ & $\Omega_{\mathrm{S},33}$\\\hline 
$a_{0}$& \SI{1.21E+01}{}& \SI{1.19E-01}{}& \SI{8.74E+00}{}& \SI{9.64E+00}{}& \SI{6.48E+00}{}& \SI{4.90E-01}{}& \SI{2.86E+00}{}& \SI{8.07E+00}{}\\\hline 
$a_{1}$& \SI{-4.23E+00}{}& \SI{2.57E-01}{}& \SI{-6.66E-01}{}& \SI{-9.10E-01}{}& \SI{-6.40E+00}{}& \SI{-2.01E-01}{}& \SI{-5.42E-01}{}& \SI{-1.03E+01}{}\\\hline 
$a_{2}$& \SI{2.35E+00}{}& \SI{-4.38E-01}{}& \SI{-2.63E+00}{}& \SI{-4.27E+00}{}& \SI{-3.07E-01}{}& \SI{-7.43E-01}{}& \SI{-4.08E-01}{}& \SI{5.70E+00}{}\\\hline 
$a_{3}$& \SI{-2.12E+00}{}& \SI{1.85E-02}{}& \SI{2.82E+00}{}& \SI{3.98E+00}{}& \SI{2.98E+00}{}& \SI{6.63E-01}{}& \SI{-1.45E-01}{}& \SI{-2.14E+00}{}\\\hline 
$\mathrm{RMSD}$& \SI{3.46E-02}{}& \SI{3.53E-03}{}& \SI{2.09E-02}{}& \SI{1.95E-02}{}& \SI{1.76E-02}{}& \SI{3.78E-03}{}& \SI{9.51E-03}{}& \SI{1.61E-02}{}\\\hline  
\end{tabular} 
\caption{\label{tab:double_cup_cones_hollow}Analogous to Tab.\ \ref{tab:half_ball}, but now for a \textit{double-cup particle consisting of two hollow right circular cones}, which have diameter $\sigma$ and wall thickness $w$ (see Fig.\ \ref{fig:double_cups}a). The used mesh included \SI{8.77E+04} to \SI{1.17E+05} triangles on the particle surface and \SI{2.31E+06} to \SI{2.57E+06} tetrahedra outside of the particle.}
\end{table*}

\clearpage
\FloatBarrier
\twocolumngrid
\bibliographystyle{apsrev4-1}
\bibliography{refs}

\begin{thebibliography}{58}%
\makeatletter
\providecommand \@ifxundefined [1]{%
 \@ifx{#1\undefined}
}%
\providecommand \@ifnum [1]{%
 \ifnum #1\expandafter \@firstoftwo
 \else \expandafter \@secondoftwo
 \fi
}%
\providecommand \@ifx [1]{%
 \ifx #1\expandafter \@firstoftwo
 \else \expandafter \@secondoftwo
 \fi
}%
\providecommand \natexlab [1]{#1}%
\providecommand \enquote  [1]{``#1''}%
\providecommand \bibnamefont  [1]{#1}%
\providecommand \bibfnamefont [1]{#1}%
\providecommand \citenamefont [1]{#1}%
\providecommand \href@noop [0]{\@secondoftwo}%
\providecommand \href [0]{\begingroup \@sanitize@url \@href}%
\providecommand \@href[1]{\@@startlink{#1}\@@href}%
\providecommand \@@href[1]{\endgroup#1\@@endlink}%
\providecommand \@sanitize@url [0]{\catcode `\\12\catcode `\$12\catcode
  `\&12\catcode `\#12\catcode `\^12\catcode `\_12\catcode `\%12\relax}%
\providecommand \@@startlink[1]{}%
\providecommand \@@endlink[0]{}%
\providecommand \url  [0]{\begingroup\@sanitize@url \@url }%
\providecommand \@url [1]{\endgroup\@href {#1}{\urlprefix }}%
\providecommand \urlprefix  [0]{URL }%
\providecommand \Eprint [0]{\href }%
\providecommand \doibase [0]{http://dx.doi.org/}%
\providecommand \selectlanguage [0]{\@gobble}%
\providecommand \bibinfo  [0]{\@secondoftwo}%
\providecommand \bibfield  [0]{\@secondoftwo}%
\providecommand \translation [1]{[#1]}%
\providecommand \BibitemOpen [0]{}%
\providecommand \bibitemStop [0]{}%
\providecommand \bibitemNoStop [0]{.\EOS\space}%
\providecommand \EOS [0]{\spacefactor3000\relax}%
\providecommand \BibitemShut  [1]{\csname bibitem#1\endcsname}%
\let\auto@bib@innerbib\@empty
\bibitem [{\citenamefont {{Yin}}\ and\ \citenamefont
  {{Alivisatos}}(2005)}]{YinA2005}%
  \BibitemOpen
  \bibfield  {author} {\bibinfo {author} {\bibfnamefont {Y.}~\bibnamefont
  {{Yin}}}\ and\ \bibinfo {author} {\bibfnamefont {A.~P.}\ \bibnamefont
  {{Alivisatos}}},\ }\href@noop {} {\bibfield  {journal} {\bibinfo  {journal}
  {Nature}\ }\textbf {\bibinfo {volume} {437}},\ \bibinfo {pages} {664}
  (\bibinfo {year} {2005})}\BibitemShut {NoStop}%
\bibitem [{\citenamefont {Burda}\ \emph {et~al.}(2005)\citenamefont {Burda},
  \citenamefont {Chen}, \citenamefont {Narayanan},\ and\ \citenamefont
  {{El-Sayed}}}]{BurdaCNES2005}%
  \BibitemOpen
  \bibfield  {author} {\bibinfo {author} {\bibfnamefont {C.}~\bibnamefont
  {Burda}}, \bibinfo {author} {\bibfnamefont {X.}~\bibnamefont {Chen}},
  \bibinfo {author} {\bibfnamefont {R.}~\bibnamefont {Narayanan}}, \ and\
  \bibinfo {author} {\bibfnamefont {M.~A.}\ \bibnamefont {{El-Sayed}}},\
  }\href@noop {} {\bibfield  {journal} {\bibinfo  {journal} {Chemical Reviews}\
  }\textbf {\bibinfo {volume} {105}},\ \bibinfo {pages} {1025} (\bibinfo {year}
  {2005})}\BibitemShut {NoStop}%
\bibitem [{\citenamefont {Tao}\ \emph {et~al.}(2008)\citenamefont {Tao},
  \citenamefont {Habas},\ and\ \citenamefont {Yang}}]{TaoHY2008}%
  \BibitemOpen
  \bibfield  {author} {\bibinfo {author} {\bibfnamefont {A.~R.}\ \bibnamefont
  {Tao}}, \bibinfo {author} {\bibfnamefont {S.}~\bibnamefont {Habas}}, \ and\
  \bibinfo {author} {\bibfnamefont {P.}~\bibnamefont {Yang}},\ }\href@noop {}
  {\bibfield  {journal} {\bibinfo  {journal} {Small}\ }\textbf {\bibinfo
  {volume} {4}},\ \bibinfo {pages} {310} (\bibinfo {year} {2008})}\BibitemShut
  {NoStop}%
\bibitem [{\citenamefont {Champion}\ \emph {et~al.}(2007)\citenamefont
  {Champion}, \citenamefont {Katare},\ and\ \citenamefont
  {Mitragotri}}]{ChampionKM2007}%
  \BibitemOpen
  \bibfield  {author} {\bibinfo {author} {\bibfnamefont {J.~A.}\ \bibnamefont
  {Champion}}, \bibinfo {author} {\bibfnamefont {Y.~K.}\ \bibnamefont
  {Katare}}, \ and\ \bibinfo {author} {\bibfnamefont {S.}~\bibnamefont
  {Mitragotri}},\ }\href@noop {} {\bibfield  {journal} {\bibinfo  {journal}
  {Journal of Controlled Release}\ }\textbf {\bibinfo {volume} {121}},\
  \bibinfo {pages} {3} (\bibinfo {year} {2007})}\BibitemShut {NoStop}%
\bibitem [{\citenamefont {Sacanna}\ and\ \citenamefont
  {Pine}(2011)}]{SacannaP2011}%
  \BibitemOpen
  \bibfield  {author} {\bibinfo {author} {\bibfnamefont {S.}~\bibnamefont
  {Sacanna}}\ and\ \bibinfo {author} {\bibfnamefont {D.~J.}\ \bibnamefont
  {Pine}},\ }\href@noop {} {\bibfield  {journal} {\bibinfo  {journal} {Current
  Opinion in Colloid \& Interface Science}\ }\textbf {\bibinfo {volume} {16}},\
  \bibinfo {pages} {96} (\bibinfo {year} {2011})}\BibitemShut {NoStop}%
\bibitem [{\citenamefont {Kuijk}\ \emph {et~al.}(2011)\citenamefont {Kuijk},
  \citenamefont {{van Blaaderen}},\ and\ \citenamefont {Imhof}}]{KuijkvBI2011}%
  \BibitemOpen
  \bibfield  {author} {\bibinfo {author} {\bibfnamefont {A.}~\bibnamefont
  {Kuijk}}, \bibinfo {author} {\bibfnamefont {A.}~\bibnamefont {{van
  Blaaderen}}}, \ and\ \bibinfo {author} {\bibfnamefont {A.}~\bibnamefont
  {Imhof}},\ }\href@noop {} {\bibfield  {journal} {\bibinfo  {journal} {Journal
  of the American Chemical Society}\ }\textbf {\bibinfo {volume} {133}},\
  \bibinfo {pages} {2346} (\bibinfo {year} {2011})}\BibitemShut {NoStop}%
\bibitem [{\citenamefont {Brenner}(1967)}]{Brenner1967}%
  \BibitemOpen
  \bibfield  {author} {\bibinfo {author} {\bibfnamefont {H.}~\bibnamefont
  {Brenner}},\ }\href@noop {} {\bibfield  {journal} {\bibinfo  {journal}
  {Journal of Colloid and Interface Science}\ }\textbf {\bibinfo {volume}
  {23}},\ \bibinfo {pages} {407} (\bibinfo {year} {1967})}\BibitemShut
  {NoStop}%
\bibitem [{\citenamefont {Happel}\ and\ \citenamefont
  {Brenner}(1991)}]{HappelB1991}%
  \BibitemOpen
  \bibfield  {author} {\bibinfo {author} {\bibfnamefont {J.}~\bibnamefont
  {Happel}}\ and\ \bibinfo {author} {\bibfnamefont {H.}~\bibnamefont
  {Brenner}},\ }\href@noop {} {\emph {\bibinfo {title} {Low {R}eynolds Number
  Hydrodynamics: With Special Applications to Particulate Media}}},\ \bibinfo
  {edition} {2nd}\ ed.,\ \bibinfo {series} {Mechanics of Fluids and Transport
  Processes}, Vol.~\bibinfo {volume} {1}\ (\bibinfo  {publisher} {Kluwer
  Academic Publishers},\ \bibinfo {address} {Dordrecht},\ \bibinfo {year}
  {1991})\BibitemShut {NoStop}%
\bibitem [{\citenamefont {Kraft}\ \emph {et~al.}(2013)\citenamefont {Kraft},
  \citenamefont {Wittkowski}, \citenamefont {{ten Hagen}}, \citenamefont
  {Edmond}, \citenamefont {Pine},\ and\ \citenamefont
  {L\"owen}}]{KraftWtHEPL2013}%
  \BibitemOpen
  \bibfield  {author} {\bibinfo {author} {\bibfnamefont {D.~J.}\ \bibnamefont
  {Kraft}}, \bibinfo {author} {\bibfnamefont {R.}~\bibnamefont {Wittkowski}},
  \bibinfo {author} {\bibfnamefont {B.}~\bibnamefont {{ten Hagen}}}, \bibinfo
  {author} {\bibfnamefont {K.~V.}\ \bibnamefont {Edmond}}, \bibinfo {author}
  {\bibfnamefont {D.~J.}\ \bibnamefont {Pine}}, \ and\ \bibinfo {author}
  {\bibfnamefont {H.}~\bibnamefont {L\"owen}},\ }\href@noop {} {\bibfield
  {journal} {\bibinfo  {journal} {Physical Review E}\ }\textbf {\bibinfo
  {volume} {88}},\ \bibinfo {pages} {050301(R)} (\bibinfo {year}
  {2013})}\BibitemShut {NoStop}%
\bibitem [{\citenamefont {Bechinger}\ \emph {et~al.}(2016)\citenamefont
  {Bechinger}, \citenamefont {{Di Leonardo}}, \citenamefont {L{\"o}wen},
  \citenamefont {Reichhardt}, \citenamefont {Volpe},\ and\ \citenamefont
  {Volpe}}]{BechingerdLLRVV2016}%
  \BibitemOpen
  \bibfield  {author} {\bibinfo {author} {\bibfnamefont {C.}~\bibnamefont
  {Bechinger}}, \bibinfo {author} {\bibfnamefont {R.}~\bibnamefont {{Di
  Leonardo}}}, \bibinfo {author} {\bibfnamefont {H.}~\bibnamefont {L{\"o}wen}},
  \bibinfo {author} {\bibfnamefont {C.}~\bibnamefont {Reichhardt}}, \bibinfo
  {author} {\bibfnamefont {G.}~\bibnamefont {Volpe}}, \ and\ \bibinfo {author}
  {\bibfnamefont {G.}~\bibnamefont {Volpe}},\ }\href@noop {} {\bibfield
  {journal} {\bibinfo  {journal} {Reviews of Modern Physics}\ }\textbf
  {\bibinfo {volume} {88}},\ \bibinfo {pages} {045006} (\bibinfo {year}
  {2016})}\BibitemShut {NoStop}%
\bibitem [{\citenamefont {Wittkowski}\ and\ \citenamefont
  {L{\"o}wen}(2012)}]{WittkowskiL2012}%
  \BibitemOpen
  \bibfield  {author} {\bibinfo {author} {\bibfnamefont {R.}~\bibnamefont
  {Wittkowski}}\ and\ \bibinfo {author} {\bibfnamefont {H.}~\bibnamefont
  {L{\"o}wen}},\ }\href@noop {} {\bibfield  {journal} {\bibinfo  {journal}
  {Physical Review E}\ }\textbf {\bibinfo {volume} {85}},\ \bibinfo {pages}
  {021406} (\bibinfo {year} {2012})}\BibitemShut {NoStop}%
\bibitem [{\citenamefont {K{\"u}mmel}\ \emph {et~al.}(2013)\citenamefont
  {K{\"u}mmel}, \citenamefont {{ten Hagen}}, \citenamefont {Wittkowski},
  \citenamefont {Buttinoni}, \citenamefont {Eichhorn}, \citenamefont {Volpe},
  \citenamefont {L{\"o}wen},\ and\ \citenamefont
  {Bechinger}}]{KuemmeltHWBEVLB2013}%
  \BibitemOpen
  \bibfield  {author} {\bibinfo {author} {\bibfnamefont {F.}~\bibnamefont
  {K{\"u}mmel}}, \bibinfo {author} {\bibfnamefont {B.}~\bibnamefont {{ten
  Hagen}}}, \bibinfo {author} {\bibfnamefont {R.}~\bibnamefont {Wittkowski}},
  \bibinfo {author} {\bibfnamefont {I.}~\bibnamefont {Buttinoni}}, \bibinfo
  {author} {\bibfnamefont {R.}~\bibnamefont {Eichhorn}}, \bibinfo {author}
  {\bibfnamefont {G.}~\bibnamefont {Volpe}}, \bibinfo {author} {\bibfnamefont
  {H.}~\bibnamefont {L{\"o}wen}}, \ and\ \bibinfo {author} {\bibfnamefont
  {C.}~\bibnamefont {Bechinger}},\ }\href@noop {} {\bibfield  {journal}
  {\bibinfo  {journal} {Physical Review Letters}\ }\textbf {\bibinfo {volume}
  {110}},\ \bibinfo {pages} {198302} (\bibinfo {year} {2013})}\BibitemShut
  {NoStop}%
\bibitem [{\citenamefont {K{\"u}mmel}\ \emph {et~al.}(2014)\citenamefont
  {K{\"u}mmel}, \citenamefont {{ten Hagen}}, \citenamefont {Wittkowski},
  \citenamefont {Takagi}, \citenamefont {Buttinoni}, \citenamefont {Eichhorn},
  \citenamefont {Volpe}, \citenamefont {L{\"o}wen},\ and\ \citenamefont
  {Bechinger}}]{KuemmeltHWTBEVLB2014}%
  \BibitemOpen
  \bibfield  {author} {\bibinfo {author} {\bibfnamefont {F.}~\bibnamefont
  {K{\"u}mmel}}, \bibinfo {author} {\bibfnamefont {B.}~\bibnamefont {{ten
  Hagen}}}, \bibinfo {author} {\bibfnamefont {R.}~\bibnamefont {Wittkowski}},
  \bibinfo {author} {\bibfnamefont {D.}~\bibnamefont {Takagi}}, \bibinfo
  {author} {\bibfnamefont {I.}~\bibnamefont {Buttinoni}}, \bibinfo {author}
  {\bibfnamefont {R.}~\bibnamefont {Eichhorn}}, \bibinfo {author}
  {\bibfnamefont {G.}~\bibnamefont {Volpe}}, \bibinfo {author} {\bibfnamefont
  {H.}~\bibnamefont {L{\"o}wen}}, \ and\ \bibinfo {author} {\bibfnamefont
  {C.}~\bibnamefont {Bechinger}},\ }\href@noop {} {\bibfield  {journal}
  {\bibinfo  {journal} {Physical Review Letters}\ }\textbf {\bibinfo {volume}
  {113}},\ \bibinfo {pages} {029802} (\bibinfo {year} {2014})}\BibitemShut
  {NoStop}%
\bibitem [{\citenamefont {{ten Hagen}}\ \emph {et~al.}(2014)\citenamefont {{ten
  Hagen}}, \citenamefont {K{\"u}mmel}, \citenamefont {Wittkowski},
  \citenamefont {Takagi}, \citenamefont {L{\"o}wen},\ and\ \citenamefont
  {Bechinger}}]{tenHagenKWTLB2014}%
  \BibitemOpen
  \bibfield  {author} {\bibinfo {author} {\bibfnamefont {B.}~\bibnamefont {{ten
  Hagen}}}, \bibinfo {author} {\bibfnamefont {F.}~\bibnamefont {K{\"u}mmel}},
  \bibinfo {author} {\bibfnamefont {R.}~\bibnamefont {Wittkowski}}, \bibinfo
  {author} {\bibfnamefont {D.}~\bibnamefont {Takagi}}, \bibinfo {author}
  {\bibfnamefont {H.}~\bibnamefont {L{\"o}wen}}, \ and\ \bibinfo {author}
  {\bibfnamefont {C.}~\bibnamefont {Bechinger}},\ }\href@noop {} {\bibfield
  {journal} {\bibinfo  {journal} {Nature Communications}\ }\textbf {\bibinfo
  {volume} {5}},\ \bibinfo {pages} {4829} (\bibinfo {year} {2014})}\BibitemShut
  {NoStop}%
\bibitem [{\citenamefont {Dhont}(1996)}]{Dhont1996}%
  \BibitemOpen
  \bibfield  {author} {\bibinfo {author} {\bibfnamefont {J.~K.~G.}\
  \bibnamefont {Dhont}},\ }\href@noop {} {\emph {\bibinfo {title} {An
  Introduction to Dynamics of Colloids}}},\ \bibinfo {edition} {1st}\ ed.,\
  \bibinfo {series} {Studies in Interface Science}, Vol.~\bibinfo {volume} {2}\
  (\bibinfo  {publisher} {Elsevier Science},\ \bibinfo {address} {Amsterdam},\
  \bibinfo {year} {1996})\BibitemShut {NoStop}%
\bibitem [{\citenamefont {Wittkowski}\ and\ \citenamefont
  {L{\"o}wen}(2011)}]{WittkowskiL2011}%
  \BibitemOpen
  \bibfield  {author} {\bibinfo {author} {\bibfnamefont {R.}~\bibnamefont
  {Wittkowski}}\ and\ \bibinfo {author} {\bibfnamefont {H.}~\bibnamefont
  {L{\"o}wen}},\ }\href@noop {} {\bibfield  {journal} {\bibinfo  {journal}
  {Molecular Physics}\ }\textbf {\bibinfo {volume} {109}},\ \bibinfo {pages}
  {2935} (\bibinfo {year} {2011})}\BibitemShut {NoStop}%
\bibitem [{\citenamefont {Perrin}(1936)}]{Perrin1936}%
  \BibitemOpen
  \bibfield  {author} {\bibinfo {author} {\bibfnamefont {F.}~\bibnamefont
  {Perrin}},\ }\href@noop {} {\bibfield  {journal} {\bibinfo  {journal}
  {Journal de Physique et Le Radium}\ }\textbf {\bibinfo {volume} {7}},\
  \bibinfo {pages} {1} (\bibinfo {year} {1936})}\BibitemShut {NoStop}%
\bibitem [{\citenamefont {Simha}(1940)}]{Simha1940}%
  \BibitemOpen
  \bibfield  {author} {\bibinfo {author} {\bibfnamefont {R.}~\bibnamefont
  {Simha}},\ }\href@noop {} {\bibfield  {journal} {\bibinfo  {journal} {Journal
  of Physical Chemistry}\ }\textbf {\bibinfo {volume} {44}},\ \bibinfo {pages}
  {25} (\bibinfo {year} {1940})}\BibitemShut {NoStop}%
\bibitem [{\citenamefont {Batchelor}(1970)}]{Batchelor1970}%
  \BibitemOpen
  \bibfield  {author} {\bibinfo {author} {\bibfnamefont {G.~K.}\ \bibnamefont
  {Batchelor}},\ }\href@noop {} {\bibfield  {journal} {\bibinfo  {journal}
  {Journal of Fluid Mechanics}\ }\textbf {\bibinfo {volume} {44}},\ \bibinfo
  {pages} {491} (\bibinfo {year} {1970})}\BibitemShut {NoStop}%
\bibitem [{\citenamefont {Fagan}\ \emph {et~al.}(2008)\citenamefont {Fagan},
  \citenamefont {Becker}, \citenamefont {Chun},\ and\ \citenamefont
  {Hobbie}}]{FaganBCH2008}%
  \BibitemOpen
  \bibfield  {author} {\bibinfo {author} {\bibfnamefont {J.~A.}\ \bibnamefont
  {Fagan}}, \bibinfo {author} {\bibfnamefont {M.~L.}\ \bibnamefont {Becker}},
  \bibinfo {author} {\bibfnamefont {J.}~\bibnamefont {Chun}}, \ and\ \bibinfo
  {author} {\bibfnamefont {E.~K.}\ \bibnamefont {Hobbie}},\ }\href@noop {}
  {\bibfield  {journal} {\bibinfo  {journal} {Advanced Materials}\ }\textbf
  {\bibinfo {volume} {20}},\ \bibinfo {pages} {1609} (\bibinfo {year}
  {2008})}\BibitemShut {NoStop}%
\bibitem [{\citenamefont {Agarwal}\ \emph {et~al.}(2005)\citenamefont
  {Agarwal}, \citenamefont {Ladavac}, \citenamefont {Roichman}, \citenamefont
  {Yu}, \citenamefont {Lieber},\ and\ \citenamefont
  {Grier}}]{AgarwalLRYLG2005}%
  \BibitemOpen
  \bibfield  {author} {\bibinfo {author} {\bibfnamefont {R.}~\bibnamefont
  {Agarwal}}, \bibinfo {author} {\bibfnamefont {K.}~\bibnamefont {Ladavac}},
  \bibinfo {author} {\bibfnamefont {Y.}~\bibnamefont {Roichman}}, \bibinfo
  {author} {\bibfnamefont {G.}~\bibnamefont {Yu}}, \bibinfo {author}
  {\bibfnamefont {C.~M.}\ \bibnamefont {Lieber}}, \ and\ \bibinfo {author}
  {\bibfnamefont {D.~G.}\ \bibnamefont {Grier}},\ }\href@noop {} {\bibfield
  {journal} {\bibinfo  {journal} {Optics Express}\ }\textbf {\bibinfo {volume}
  {13}},\ \bibinfo {pages} {8906} (\bibinfo {year} {2005})}\BibitemShut
  {NoStop}%
\bibitem [{\citenamefont {{ten Hagen}}\ \emph {et~al.}(2015)\citenamefont {{ten
  Hagen}}, \citenamefont {Wittkowski}, \citenamefont {Takagi}, \citenamefont
  {K{\"u}mmel}, \citenamefont {Bechinger},\ and\ \citenamefont
  {L{\"o}wen}}]{tenHagenWTKBL2015}%
  \BibitemOpen
  \bibfield  {author} {\bibinfo {author} {\bibfnamefont {B.}~\bibnamefont {{ten
  Hagen}}}, \bibinfo {author} {\bibfnamefont {R.}~\bibnamefont {Wittkowski}},
  \bibinfo {author} {\bibfnamefont {D.}~\bibnamefont {Takagi}}, \bibinfo
  {author} {\bibfnamefont {F.}~\bibnamefont {K{\"u}mmel}}, \bibinfo {author}
  {\bibfnamefont {C.}~\bibnamefont {Bechinger}}, \ and\ \bibinfo {author}
  {\bibfnamefont {H.}~\bibnamefont {L{\"o}wen}},\ }\href@noop {} {\bibfield
  {journal} {\bibinfo  {journal} {Journal of Physics: Condensed Matter}\
  }\textbf {\bibinfo {volume} {27}},\ \bibinfo {pages} {194110} (\bibinfo
  {year} {2015})}\BibitemShut {NoStop}%
\bibitem [{\citenamefont {Clague}\ and\ \citenamefont
  {Phillips}(1997)}]{ClagueP1997}%
  \BibitemOpen
  \bibfield  {author} {\bibinfo {author} {\bibfnamefont {D.~S.}\ \bibnamefont
  {Clague}}\ and\ \bibinfo {author} {\bibfnamefont {R.~J.}\ \bibnamefont
  {Phillips}},\ }\href@noop {} {\bibfield  {journal} {\bibinfo  {journal}
  {Physics of Fluids}\ }\textbf {\bibinfo {volume} {9}},\ \bibinfo {pages}
  {1562} (\bibinfo {year} {1997})}\BibitemShut {NoStop}%
\bibitem [{\citenamefont {Switzer}\ and\ \citenamefont
  {Klingenberg}(2003)}]{SwitzerK2003}%
  \BibitemOpen
  \bibfield  {author} {\bibinfo {author} {\bibfnamefont {L.~H.}\ \bibnamefont
  {Switzer}}\ and\ \bibinfo {author} {\bibfnamefont {D.~J.}\ \bibnamefont
  {Klingenberg}},\ }\href@noop {} {\bibfield  {journal} {\bibinfo  {journal}
  {Journal of Rheology}\ }\textbf {\bibinfo {volume} {47}},\ \bibinfo {pages}
  {759} (\bibinfo {year} {2003})}\BibitemShut {NoStop}%
\bibitem [{\citenamefont {Swanson}\ \emph {et~al.}(1978)\citenamefont
  {Swanson}, \citenamefont {Teller},\ and\ \citenamefont {{de
  Ha\"en}}}]{SwansonTdH1978}%
  \BibitemOpen
  \bibfield  {author} {\bibinfo {author} {\bibfnamefont {E.}~\bibnamefont
  {Swanson}}, \bibinfo {author} {\bibfnamefont {D.~C.}\ \bibnamefont {Teller}},
  \ and\ \bibinfo {author} {\bibfnamefont {C.}~\bibnamefont {{de Ha\"en}}},\
  }\href@noop {} {\bibfield  {journal} {\bibinfo  {journal} {Journal of
  Chemical Physics}\ }\textbf {\bibinfo {volume} {68}},\ \bibinfo {pages}
  {5097} (\bibinfo {year} {1978})}\BibitemShut {NoStop}%
\bibitem [{\citenamefont {{Garc\'{\i}a de la Torre}}\ and\ \citenamefont
  {{Bloomfield}}(1981)}]{GarciadelaTorreB1981}%
  \BibitemOpen
  \bibfield  {author} {\bibinfo {author} {\bibfnamefont {J.}~\bibnamefont
  {{Garc\'{\i}a de la Torre}}}\ and\ \bibinfo {author} {\bibfnamefont {V.~A.}\
  \bibnamefont {{Bloomfield}}},\ }\href@noop {} {\bibfield  {journal} {\bibinfo
   {journal} {Quarterly Reviews of Biophysics}\ }\textbf {\bibinfo {volume}
  {14}},\ \bibinfo {pages} {81} (\bibinfo {year} {1981})}\BibitemShut {NoStop}%
\bibitem [{\citenamefont {{Carrasco}}\ and\ \citenamefont {{Garc\'{\i}a de la
  Torre}}(1999)}]{CarrascoGdlT1999}%
  \BibitemOpen
  \bibfield  {author} {\bibinfo {author} {\bibfnamefont {B.}~\bibnamefont
  {{Carrasco}}}\ and\ \bibinfo {author} {\bibfnamefont {J.}~\bibnamefont
  {{Garc\'{\i}a de la Torre}}},\ }\href@noop {} {\bibfield  {journal} {\bibinfo
   {journal} {Biophysical Journal}\ }\textbf {\bibinfo {volume} {76}},\
  \bibinfo {pages} {3044} (\bibinfo {year} {1999})}\BibitemShut {NoStop}%
\bibitem [{\citenamefont {{Garc\'{\i}a de la Torre}}\ and\ \citenamefont
  {{Carrasco}}(2002)}]{GarciadelaTorreC2002}%
  \BibitemOpen
  \bibfield  {author} {\bibinfo {author} {\bibfnamefont {J.}~\bibnamefont
  {{Garc\'{\i}a de la Torre}}}\ and\ \bibinfo {author} {\bibfnamefont
  {B.}~\bibnamefont {{Carrasco}}},\ }\href@noop {} {\bibfield  {journal}
  {\bibinfo  {journal} {Biopolymers}\ }\textbf {\bibinfo {volume} {63}},\
  \bibinfo {pages} {163} (\bibinfo {year} {2002})}\BibitemShut {NoStop}%
\bibitem [{\citenamefont {Hansen}(2004)}]{Hansen2004}%
  \BibitemOpen
  \bibfield  {author} {\bibinfo {author} {\bibfnamefont {S.}~\bibnamefont
  {Hansen}},\ }\href@noop {} {\bibfield  {journal} {\bibinfo  {journal}
  {Journal of Chemical Physics}\ }\textbf {\bibinfo {volume} {121}},\ \bibinfo
  {pages} {9111} (\bibinfo {year} {2004})}\BibitemShut {NoStop}%
\bibitem [{\citenamefont {Bet}\ \emph {et~al.}(2017)\citenamefont {Bet},
  \citenamefont {Boosten}, \citenamefont {Dijkstra},\ and\ \citenamefont {{van
  Roij}}}]{BetBDvR2017}%
  \BibitemOpen
  \bibfield  {author} {\bibinfo {author} {\bibfnamefont {B.}~\bibnamefont
  {Bet}}, \bibinfo {author} {\bibfnamefont {G.}~\bibnamefont {Boosten}},
  \bibinfo {author} {\bibfnamefont {M.}~\bibnamefont {Dijkstra}}, \ and\
  \bibinfo {author} {\bibfnamefont {R.}~\bibnamefont {{van Roij}}},\
  }\href@noop {} {\bibfield  {journal} {\bibinfo  {journal} {Journal of
  Chemical Physics}\ }\textbf {\bibinfo {volume} {146}},\ \bibinfo {pages}
  {084904} (\bibinfo {year} {2017})}\BibitemShut {NoStop}%
\bibitem [{\citenamefont {Tirado}\ and\ \citenamefont {{Garc\'{\i}a de la
  Torre}}(1979)}]{TiradoGdlT1979}%
  \BibitemOpen
  \bibfield  {author} {\bibinfo {author} {\bibfnamefont {M.~M.}\ \bibnamefont
  {Tirado}}\ and\ \bibinfo {author} {\bibfnamefont {J.}~\bibnamefont
  {{Garc\'{\i}a de la Torre}}},\ }\href@noop {} {\bibfield  {journal} {\bibinfo
   {journal} {Journal of Chemical Physics}\ }\textbf {\bibinfo {volume} {71}},\
  \bibinfo {pages} {2581} (\bibinfo {year} {1979})}\BibitemShut {NoStop}%
\bibitem [{\citenamefont {Passow}\ \emph {et~al.}(2015)\citenamefont {Passow},
  \citenamefont {{ten Hagen}}, \citenamefont {L\"owen},\ and\ \citenamefont
  {Wagner}}]{PassowtHLW2015}%
  \BibitemOpen
  \bibfield  {author} {\bibinfo {author} {\bibfnamefont {C.}~\bibnamefont
  {Passow}}, \bibinfo {author} {\bibfnamefont {B.}~\bibnamefont {{ten Hagen}}},
  \bibinfo {author} {\bibfnamefont {H.}~\bibnamefont {L\"owen}}, \ and\
  \bibinfo {author} {\bibfnamefont {J.}~\bibnamefont {Wagner}},\ }\href@noop {}
  {\bibfield  {journal} {\bibinfo  {journal} {Journal of Chemical Physics}\
  }\textbf {\bibinfo {volume} {143}},\ \bibinfo {pages} {044903} (\bibinfo
  {year} {2015})}\BibitemShut {NoStop}%
\bibitem [{\citenamefont {{Garc\'{\i}a de la Torre}}\ \emph
  {et~al.}(2007)\citenamefont {{Garc\'{\i}a de la Torre}}, \citenamefont {{del
  Rio Echenique}},\ and\ \citenamefont {Ortega}}]{GarciadelaTorreREO2007}%
  \BibitemOpen
  \bibfield  {author} {\bibinfo {author} {\bibfnamefont {J.}~\bibnamefont
  {{Garc\'{\i}a de la Torre}}}, \bibinfo {author} {\bibfnamefont
  {G.}~\bibnamefont {{del Rio Echenique}}}, \ and\ \bibinfo {author}
  {\bibfnamefont {A.}~\bibnamefont {Ortega}},\ }\href@noop {} {\bibfield
  {journal} {\bibinfo  {journal} {Journal of Physical Chemistry B}\ }\textbf
  {\bibinfo {volume} {111}},\ \bibinfo {pages} {955} (\bibinfo {year}
  {2007})}\BibitemShut {NoStop}%
\bibitem [{\citenamefont {Mauer}\ \emph {et~al.}(2017)\citenamefont {Mauer},
  \citenamefont {Peltom\"aki}, \citenamefont {Poblete}, \citenamefont
  {Gompper},\ and\ \citenamefont {Fedosov}}]{MauerPPGF2017}%
  \BibitemOpen
  \bibfield  {author} {\bibinfo {author} {\bibfnamefont {J.}~\bibnamefont
  {Mauer}}, \bibinfo {author} {\bibfnamefont {M.}~\bibnamefont {Peltom\"aki}},
  \bibinfo {author} {\bibfnamefont {S.}~\bibnamefont {Poblete}}, \bibinfo
  {author} {\bibfnamefont {G.}~\bibnamefont {Gompper}}, \ and\ \bibinfo
  {author} {\bibfnamefont {D.~A.}\ \bibnamefont {Fedosov}},\ }\href@noop {}
  {\bibfield  {journal} {\bibinfo  {journal} {PLOS ONE}\ }\textbf {\bibinfo
  {volume} {12}},\ \bibinfo {pages} {e0176799} (\bibinfo {year}
  {2017})}\BibitemShut {NoStop}%
\bibitem [{\citenamefont {{Kaiser}}\ \emph {et~al.}(2014)\citenamefont
  {{Kaiser}}, \citenamefont {{Peshkov}}, \citenamefont {{Sokolov}},
  \citenamefont {{ten Hagen}}, \citenamefont {{L{\"o}wen}},\ and\ \citenamefont
  {{Aranson}}}]{KaiserPStHLA2014}%
  \BibitemOpen
  \bibfield  {author} {\bibinfo {author} {\bibfnamefont {A.}~\bibnamefont
  {{Kaiser}}}, \bibinfo {author} {\bibfnamefont {A.}~\bibnamefont {{Peshkov}}},
  \bibinfo {author} {\bibfnamefont {A.}~\bibnamefont {{Sokolov}}}, \bibinfo
  {author} {\bibfnamefont {B.}~\bibnamefont {{ten Hagen}}}, \bibinfo {author}
  {\bibfnamefont {H.}~\bibnamefont {{L{\"o}wen}}}, \ and\ \bibinfo {author}
  {\bibfnamefont {I.~S.}\ \bibnamefont {{Aranson}}},\ }\href@noop {} {\bibfield
   {journal} {\bibinfo  {journal} {Physical Review Letters}\ }\textbf {\bibinfo
  {volume} {112}},\ \bibinfo {pages} {158101} (\bibinfo {year}
  {2014})}\BibitemShut {NoStop}%
\bibitem [{\citenamefont {Jeffrey}\ and\ \citenamefont
  {Onishi}(1984)}]{JeffreyO1984}%
  \BibitemOpen
  \bibfield  {author} {\bibinfo {author} {\bibfnamefont {D.~J.}\ \bibnamefont
  {Jeffrey}}\ and\ \bibinfo {author} {\bibfnamefont {Y.}~\bibnamefont
  {Onishi}},\ }\href@noop {} {\bibfield  {journal} {\bibinfo  {journal}
  {Journal of Fluid Mechanics}\ }\textbf {\bibinfo {volume} {139}},\ \bibinfo
  {pages} {261} (\bibinfo {year} {1984})}\BibitemShut {NoStop}%
\bibitem [{\citenamefont {Carlson}\ \emph {et~al.}(2006)\citenamefont
  {Carlson}, \citenamefont {Jena}, \citenamefont {Flenniken}, \citenamefont
  {Chou}, \citenamefont {Siegel},\ and\ \citenamefont
  {Wagner}}]{CarlsonJFCSW2006}%
  \BibitemOpen
  \bibfield  {author} {\bibinfo {author} {\bibfnamefont {J.~C.~T.}\
  \bibnamefont {Carlson}}, \bibinfo {author} {\bibfnamefont {S.~S.}\
  \bibnamefont {Jena}}, \bibinfo {author} {\bibfnamefont {M.}~\bibnamefont
  {Flenniken}}, \bibinfo {author} {\bibfnamefont {T.}~\bibnamefont {Chou}},
  \bibinfo {author} {\bibfnamefont {R.~A.}\ \bibnamefont {Siegel}}, \ and\
  \bibinfo {author} {\bibfnamefont {C.~R.}\ \bibnamefont {Wagner}},\
  }\href@noop {} {\bibfield  {journal} {\bibinfo  {journal} {Journal of the
  American Chemical Society}\ }\textbf {\bibinfo {volume} {128}},\ \bibinfo
  {pages} {7630} (\bibinfo {year} {2006})}\BibitemShut {NoStop}%
\bibitem [{\citenamefont {{Bernal Garcia}}\ and\ \citenamefont {{Garc\'{\i}a de
  la Torre}}(1980)}]{BernalGarciaGdlT1980}%
  \BibitemOpen
  \bibfield  {author} {\bibinfo {author} {\bibfnamefont {J.~M.}\ \bibnamefont
  {{Bernal Garcia}}}\ and\ \bibinfo {author} {\bibfnamefont {J.}~\bibnamefont
  {{Garc\'{\i}a de la Torre}}},\ }\href@noop {} {\bibfield  {journal} {\bibinfo
   {journal} {Biopolymers}\ }\textbf {\bibinfo {volume} {19}},\ \bibinfo
  {pages} {751} (\bibinfo {year} {1980})}\BibitemShut {NoStop}%
\bibitem [{\citenamefont {Matulis}\ \emph {et~al.}(1999)\citenamefont
  {Matulis}, \citenamefont {Baumann}, \citenamefont {Bloomfield},\ and\
  \citenamefont {Lovrien}}]{MatulisBBL1999}%
  \BibitemOpen
  \bibfield  {author} {\bibinfo {author} {\bibfnamefont {D.}~\bibnamefont
  {Matulis}}, \bibinfo {author} {\bibfnamefont {C.~G.}\ \bibnamefont
  {Baumann}}, \bibinfo {author} {\bibfnamefont {V.~A.}\ \bibnamefont
  {Bloomfield}}, \ and\ \bibinfo {author} {\bibfnamefont {R.~E.}\ \bibnamefont
  {Lovrien}},\ }\href@noop {} {\bibfield  {journal} {\bibinfo  {journal}
  {Biopolymers}\ }\textbf {\bibinfo {volume} {49}},\ \bibinfo {pages} {451}
  (\bibinfo {year} {1999})}\BibitemShut {NoStop}%
\bibitem [{\citenamefont {{Garc\'{\i}a de la Torre}}\ \emph
  {et~al.}(2000{\natexlab{a}})\citenamefont {{Garc\'{\i}a de la Torre}},
  \citenamefont {Huertas},\ and\ \citenamefont
  {Carrasco}}]{GarciadelaTorreHC2000a}%
  \BibitemOpen
  \bibfield  {author} {\bibinfo {author} {\bibfnamefont {J.}~\bibnamefont
  {{Garc\'{\i}a de la Torre}}}, \bibinfo {author} {\bibfnamefont {M.~L.}\
  \bibnamefont {Huertas}}, \ and\ \bibinfo {author} {\bibfnamefont
  {B.}~\bibnamefont {Carrasco}},\ }\href@noop {} {\bibfield  {journal}
  {\bibinfo  {journal} {Biophysical Journal}\ }\textbf {\bibinfo {volume}
  {78}},\ \bibinfo {pages} {719} (\bibinfo {year}
  {2000}{\natexlab{a}})}\BibitemShut {NoStop}%
\bibitem [{\citenamefont {{Garc\'{\i}a de la Torre}}\ \emph
  {et~al.}(2000{\natexlab{b}})\citenamefont {{Garc\'{\i}a de la Torre}},
  \citenamefont {Huertas},\ and\ \citenamefont
  {Carrasco}}]{GarciadelaTorreHC2000b}%
  \BibitemOpen
  \bibfield  {author} {\bibinfo {author} {\bibfnamefont {J.}~\bibnamefont
  {{Garc\'{\i}a de la Torre}}}, \bibinfo {author} {\bibfnamefont {M.~L.}\
  \bibnamefont {Huertas}}, \ and\ \bibinfo {author} {\bibfnamefont
  {B.}~\bibnamefont {Carrasco}},\ }\href@noop {} {\bibfield  {journal}
  {\bibinfo  {journal} {Journal of Magnetic Resonance}\ }\textbf {\bibinfo
  {volume} {147}},\ \bibinfo {pages} {138} (\bibinfo {year}
  {2000}{\natexlab{b}})}\BibitemShut {NoStop}%
\bibitem [{\citenamefont {Fung}\ and\ \citenamefont
  {Manoharan}(2013)}]{FungM2013}%
  \BibitemOpen
  \bibfield  {author} {\bibinfo {author} {\bibfnamefont {J.}~\bibnamefont
  {Fung}}\ and\ \bibinfo {author} {\bibfnamefont {V.~N.}\ \bibnamefont
  {Manoharan}},\ }\href@noop {} {\bibfield  {journal} {\bibinfo  {journal}
  {Physical Review E}\ }\textbf {\bibinfo {volume} {88}},\ \bibinfo {pages}
  {020302} (\bibinfo {year} {2013})}\BibitemShut {NoStop}%
\bibitem [{\citenamefont {Wang}\ \emph {et~al.}(2012)\citenamefont {Wang},
  \citenamefont {Castro}, \citenamefont {Hoyos},\ and\ \citenamefont
  {Mallouk}}]{WangCHM2012}%
  \BibitemOpen
  \bibfield  {author} {\bibinfo {author} {\bibfnamefont {W.}~\bibnamefont
  {Wang}}, \bibinfo {author} {\bibfnamefont {L.}~\bibnamefont {Castro}},
  \bibinfo {author} {\bibfnamefont {M.}~\bibnamefont {Hoyos}}, \ and\ \bibinfo
  {author} {\bibfnamefont {T.~E.}\ \bibnamefont {Mallouk}},\ }\href@noop {}
  {\bibfield  {journal} {\bibinfo  {journal} {ACS Nano}\ }\textbf {\bibinfo
  {volume} {6}},\ \bibinfo {pages} {6122} (\bibinfo {year} {2012})}\BibitemShut
  {NoStop}%
\bibitem [{\citenamefont {{Garcia-Gradilla}}\ \emph {et~al.}(2013)\citenamefont
  {{Garcia-Gradilla}}, \citenamefont {Orozco}, \citenamefont
  {Sattayasamitsathit}, \citenamefont {Soto}, \citenamefont {Kuralay},
  \citenamefont {Pourazary}, \citenamefont {Katzenberg}, \citenamefont {Gao},
  \citenamefont {Shen},\ and\ \citenamefont {Wang}}]{GarciaGradillaEtAl2013}%
  \BibitemOpen
  \bibfield  {author} {\bibinfo {author} {\bibfnamefont {V.}~\bibnamefont
  {{Garcia-Gradilla}}}, \bibinfo {author} {\bibfnamefont {J.}~\bibnamefont
  {Orozco}}, \bibinfo {author} {\bibfnamefont {S.}~\bibnamefont
  {Sattayasamitsathit}}, \bibinfo {author} {\bibfnamefont {F.}~\bibnamefont
  {Soto}}, \bibinfo {author} {\bibfnamefont {F.}~\bibnamefont {Kuralay}},
  \bibinfo {author} {\bibfnamefont {A.}~\bibnamefont {Pourazary}}, \bibinfo
  {author} {\bibfnamefont {A.}~\bibnamefont {Katzenberg}}, \bibinfo {author}
  {\bibfnamefont {W.}~\bibnamefont {Gao}}, \bibinfo {author} {\bibfnamefont
  {Y.}~\bibnamefont {Shen}}, \ and\ \bibinfo {author} {\bibfnamefont
  {J.}~\bibnamefont {Wang}},\ }\href@noop {} {\bibfield  {journal} {\bibinfo
  {journal} {ACS Nano}\ }\textbf {\bibinfo {volume} {7}},\ \bibinfo {pages}
  {9232} (\bibinfo {year} {2013})}\BibitemShut {NoStop}%
\bibitem [{\citenamefont {Zerrouki}\ \emph {et~al.}(2008)\citenamefont
  {Zerrouki}, \citenamefont {Baudry}, \citenamefont {Pine}, \citenamefont
  {Chaikin},\ and\ \citenamefont {Bibette}}]{ZerroukiBPCB2008}%
  \BibitemOpen
  \bibfield  {author} {\bibinfo {author} {\bibfnamefont {D.}~\bibnamefont
  {Zerrouki}}, \bibinfo {author} {\bibfnamefont {J.}~\bibnamefont {Baudry}},
  \bibinfo {author} {\bibfnamefont {D.}~\bibnamefont {Pine}}, \bibinfo {author}
  {\bibfnamefont {P.}~\bibnamefont {Chaikin}}, \ and\ \bibinfo {author}
  {\bibfnamefont {J.}~\bibnamefont {Bibette}},\ }\href@noop {} {\bibfield
  {journal} {\bibinfo  {journal} {Nature}\ }\textbf {\bibinfo {volume} {455}},\
  \bibinfo {pages} {380} (\bibinfo {year} {2008})}\BibitemShut {NoStop}%
\bibitem [{\citenamefont {Sokolov}\ \emph {et~al.}(2010)\citenamefont
  {Sokolov}, \citenamefont {Apodaca}, \citenamefont {Grzybowski},\ and\
  \citenamefont {Aranson}}]{SokolovAGA2010}%
  \BibitemOpen
  \bibfield  {author} {\bibinfo {author} {\bibfnamefont {A.}~\bibnamefont
  {Sokolov}}, \bibinfo {author} {\bibfnamefont {M.~M.}\ \bibnamefont
  {Apodaca}}, \bibinfo {author} {\bibfnamefont {B.~A.}\ \bibnamefont
  {Grzybowski}}, \ and\ \bibinfo {author} {\bibfnamefont {I.~S.}\ \bibnamefont
  {Aranson}},\ }\href@noop {} {\bibfield  {journal} {\bibinfo  {journal}
  {Proceedings of the National Academy of Sciences U.S.A.}\ }\textbf {\bibinfo
  {volume} {107}},\ \bibinfo {pages} {969} (\bibinfo {year}
  {2010})}\BibitemShut {NoStop}%
\bibitem [{\citenamefont {Mijalkov}\ and\ \citenamefont
  {Volpe}(2013)}]{MijalkovV2013}%
  \BibitemOpen
  \bibfield  {author} {\bibinfo {author} {\bibfnamefont {M.}~\bibnamefont
  {Mijalkov}}\ and\ \bibinfo {author} {\bibfnamefont {G.}~\bibnamefont
  {Volpe}},\ }\href@noop {} {\bibfield  {journal} {\bibinfo  {journal} {Soft
  Matter}\ }\textbf {\bibinfo {volume} {9}},\ \bibinfo {pages} {6376} (\bibinfo
  {year} {2013})}\BibitemShut {NoStop}%
\bibitem [{\citenamefont {Schamel}\ \emph {et~al.}(2013)\citenamefont
  {Schamel}, \citenamefont {Pfeifer}, \citenamefont {Gibbs}, \citenamefont
  {Miksch}, \citenamefont {Mark},\ and\ \citenamefont
  {Fischer}}]{SchamelPGMMF2013}%
  \BibitemOpen
  \bibfield  {author} {\bibinfo {author} {\bibfnamefont {D.}~\bibnamefont
  {Schamel}}, \bibinfo {author} {\bibfnamefont {M.}~\bibnamefont {Pfeifer}},
  \bibinfo {author} {\bibfnamefont {J.~G.}\ \bibnamefont {Gibbs}}, \bibinfo
  {author} {\bibfnamefont {B.}~\bibnamefont {Miksch}}, \bibinfo {author}
  {\bibfnamefont {A.~G.}\ \bibnamefont {Mark}}, \ and\ \bibinfo {author}
  {\bibfnamefont {P.}~\bibnamefont {Fischer}},\ }\href@noop {} {\bibfield
  {journal} {\bibinfo  {journal} {Journal of the American Chemical Society}\
  }\textbf {\bibinfo {volume} {135}},\ \bibinfo {pages} {12353} (\bibinfo
  {year} {2013})}\BibitemShut {NoStop}%
\bibitem [{\citenamefont {{Yuan}}\ \emph {et~al.}(2018)\citenamefont {{Yuan}},
  \citenamefont {{Martinez}}, \citenamefont {{Senyuk}}, \citenamefont
  {{Tasinkevych}},\ and\ \citenamefont {{Smalyukh}}}]{YuanMSTS2018}%
  \BibitemOpen
  \bibfield  {author} {\bibinfo {author} {\bibfnamefont {Y.}~\bibnamefont
  {{Yuan}}}, \bibinfo {author} {\bibfnamefont {A.}~\bibnamefont {{Martinez}}},
  \bibinfo {author} {\bibfnamefont {B.}~\bibnamefont {{Senyuk}}}, \bibinfo
  {author} {\bibfnamefont {M.}~\bibnamefont {{Tasinkevych}}}, \ and\ \bibinfo
  {author} {\bibfnamefont {I.~I.}\ \bibnamefont {{Smalyukh}}},\ }\href@noop {}
  {\bibfield  {journal} {\bibinfo  {journal} {Nature Materials}\ }\textbf
  {\bibinfo {volume} {17}},\ \bibinfo {pages} {71} (\bibinfo {year}
  {2018})}\BibitemShut {NoStop}%
\bibitem [{\citenamefont {Kirby}(2004)}]{Kirby2004}%
  \BibitemOpen
  \bibfield  {author} {\bibinfo {author} {\bibfnamefont {R.~C.}\ \bibnamefont
  {Kirby}},\ }\href@noop {} {\bibfield  {journal} {\bibinfo  {journal} {ACM
  Transactions on Mathematical Software}\ }\textbf {\bibinfo {volume} {30}},\
  \bibinfo {pages} {502} (\bibinfo {year} {2004})}\BibitemShut {NoStop}%
\bibitem [{\citenamefont {Kirby}\ and\ \citenamefont
  {Logg}(2006)}]{KirbyL2006}%
  \BibitemOpen
  \bibfield  {author} {\bibinfo {author} {\bibfnamefont {R.~C.}\ \bibnamefont
  {Kirby}}\ and\ \bibinfo {author} {\bibfnamefont {A.}~\bibnamefont {Logg}},\
  }\href@noop {} {\bibfield  {journal} {\bibinfo  {journal} {ACM Transactions
  on Mathematical Software}\ }\textbf {\bibinfo {volume} {32}},\ \bibinfo
  {pages} {417} (\bibinfo {year} {2006})}\BibitemShut {NoStop}%
\bibitem [{\citenamefont {Aln\ae{}s}\ \emph {et~al.}(2009)\citenamefont
  {Aln\ae{}s}, \citenamefont {Logg}, \citenamefont {Mardal}, \citenamefont
  {Skavhaug},\ and\ \citenamefont {Langtangen}}]{AlnaesLMSL2009}%
  \BibitemOpen
  \bibfield  {author} {\bibinfo {author} {\bibfnamefont {M.~S.}\ \bibnamefont
  {Aln\ae{}s}}, \bibinfo {author} {\bibfnamefont {A.}~\bibnamefont {Logg}},
  \bibinfo {author} {\bibfnamefont {K.~A.}\ \bibnamefont {Mardal}}, \bibinfo
  {author} {\bibfnamefont {O.}~\bibnamefont {Skavhaug}}, \ and\ \bibinfo
  {author} {\bibfnamefont {H.~P.}\ \bibnamefont {Langtangen}},\ }\href@noop {}
  {\bibfield  {journal} {\bibinfo  {journal} {International Journal of
  Computational Science and Engineering}\ }\textbf {\bibinfo {volume} {4}},\
  \bibinfo {pages} {231} (\bibinfo {year} {2009})}\BibitemShut {NoStop}%
\bibitem [{\citenamefont {Logg}\ and\ \citenamefont {Wells}(2010)}]{LoggW2010}%
  \BibitemOpen
  \bibfield  {author} {\bibinfo {author} {\bibfnamefont {A.}~\bibnamefont
  {Logg}}\ and\ \bibinfo {author} {\bibfnamefont {G.~N.}\ \bibnamefont
  {Wells}},\ }\href@noop {} {\bibfield  {journal} {\bibinfo  {journal} {ACM
  Transactions on Mathematical Software}\ }\textbf {\bibinfo {volume} {37}},\
  \bibinfo {pages} {20} (\bibinfo {year} {2010})}\BibitemShut {NoStop}%
\bibitem [{\citenamefont {\O{}lgaard}\ and\ \citenamefont
  {Wells}(2010)}]{OlgaardW2010}%
  \BibitemOpen
  \bibfield  {author} {\bibinfo {author} {\bibfnamefont {K.~B.}\ \bibnamefont
  {\O{}lgaard}}\ and\ \bibinfo {author} {\bibfnamefont {G.~N.}\ \bibnamefont
  {Wells}},\ }\href@noop {} {\bibfield  {journal} {\bibinfo  {journal} {ACM
  Transactions on Mathematical Software}\ }\textbf {\bibinfo {volume} {37}},\
  \bibinfo {pages} {8} (\bibinfo {year} {2010})}\BibitemShut {NoStop}%
\bibitem [{\citenamefont {{Aln\ae{}s}}\ \emph {et~al.}(2014)\citenamefont
  {{Aln\ae{}s}}, \citenamefont {Logg}, \citenamefont {\O{}lgaard},
  \citenamefont {Rognes},\ and\ \citenamefont {Wells}}]{AlnaesLORW2012}%
  \BibitemOpen
  \bibfield  {author} {\bibinfo {author} {\bibfnamefont {M.~S.}\ \bibnamefont
  {{Aln\ae{}s}}}, \bibinfo {author} {\bibfnamefont {A.}~\bibnamefont {Logg}},
  \bibinfo {author} {\bibfnamefont {K.~B.}\ \bibnamefont {\O{}lgaard}},
  \bibinfo {author} {\bibfnamefont {M.~E.}\ \bibnamefont {Rognes}}, \ and\
  \bibinfo {author} {\bibfnamefont {G.~N.}\ \bibnamefont {Wells}},\ }\href@noop
  {} {\bibfield  {journal} {\bibinfo  {journal} {ACM Transactions on
  Mathematical Software}\ }\textbf {\bibinfo {volume} {40}},\ \bibinfo {pages}
  {9} (\bibinfo {year} {2014})}\BibitemShut {NoStop}%
\bibitem [{\citenamefont {Logg}\ \emph {et~al.}(2012)\citenamefont {Logg},
  \citenamefont {Mardal},\ and\ \citenamefont {Wells}}]{LoggMW2012}%
  \BibitemOpen
  \bibfield  {author} {\bibinfo {author} {\bibfnamefont {A.}~\bibnamefont
  {Logg}}, \bibinfo {author} {\bibfnamefont {K.-A.}\ \bibnamefont {Mardal}}, \
  and\ \bibinfo {author} {\bibfnamefont {G.~N.}\ \bibnamefont {Wells}},\
  }\href@noop {} {\emph {\bibinfo {title} {Automated Solution of Differential
  Equations by the Finite Element Method: The FEniCS Book}}},\ \bibinfo
  {edition} {1st}\ ed.,\ \bibinfo {series} {Lecture Notes in Computational
  Science and Engineering}, Vol.~\bibinfo {volume} {84}\ (\bibinfo  {publisher}
  {Springer-Verlag},\ \bibinfo {address} {Heidelberg},\ \bibinfo {year}
  {2012})\BibitemShut {NoStop}%
\bibitem [{\citenamefont {Aln{\ae}s}\ \emph {et~al.}(2015)\citenamefont
  {Aln{\ae}s} \emph {et~al.}}]{AlnaesEtAl2015}%
  \BibitemOpen
  \bibfield  {author} {\bibinfo {author} {\bibfnamefont {M.~S.}\ \bibnamefont
  {Aln{\ae}s}} \emph {et~al.},\ }\href@noop {} {\bibfield  {journal} {\bibinfo
  {journal} {Archive of Numerical Software}\ }\textbf {\bibinfo {volume} {3}}
  (\bibinfo {year} {2015})}\BibitemShut {NoStop}%
\bibitem [{\citenamefont {Geuzaine}\ and\ \citenamefont
  {Remacle}(2009)}]{GeuzaineR2009}%
  \BibitemOpen
  \bibfield  {author} {\bibinfo {author} {\bibfnamefont {C.}~\bibnamefont
  {Geuzaine}}\ and\ \bibinfo {author} {\bibfnamefont {J.~F.}\ \bibnamefont
  {Remacle}},\ }\href@noop {} {\bibfield  {journal} {\bibinfo  {journal}
  {International Journal for Numerical Methods in Engineering}\ }\textbf
  {\bibinfo {volume} {79}},\ \bibinfo {pages} {1309} (\bibinfo {year}
  {2009})}\BibitemShut {NoStop}%
\end{thebibliography}%

\end{document}